\documentclass[preprint2]{aastex}

\newcommand{\etal}{{\it et al.}}

\slugcomment{To appear in Ap. J.}

\shorttitle{EVPA reversals in SiO maser emission}
\shortauthors{Kemball et al.}

\usepackage{amssymb}

\begin{document}


\title{Electric vector rotations of $\frac{\pi}{2}$ in polarized circumstellar SiO maser emission}


\author{A. J. Kemball}
\affil{Department of Astronomy\altaffilmark{1},\\ University of Illinois at Urbana-Champaign,\\ 1002 W. Green Street, Urbana, IL 61801}
\email{akemball@illinois.edu}

\author{P. J. Diamond}
\affil{CSIRO Astronomy and Space Science,\\ Vimiera and Pembroke Roads, Marsfield NSW 2122, Australia}

\author{L. Richter}
\affil{Department of Physics and Electronics,\\ Rhodes University, P.O. Box 94, Grahamstown 6140, South Africa}

\author{I. Gonidakis}
\affil{CSIRO Astronomy and Space Science,\\ Vimiera and Pembroke Roads, Marsfield NSW 2122, Australia}

\and

\author{R. Xue}
\affil{Department of Astronomy,\\ University of Illinois at Urbana-Champaign,\\ 1002 W. Green Street, Urbana, IL 61801}


\altaffiltext{1}{Institute for Advanced Computing Applications and Technologies, University of Illinois at Urbana-Champaign}


\begin{abstract}
This paper examines the detailed sub-milliarcsecond polarization
properties of an individual SiO\ maser feature displaying a rotation
in polarization electric vector position angle of approximately
$\frac{\pi}{2}$ across the feature. Such rotations are a
characteristic observational signature of circumstellar SiO masers
detected toward a number of late-type, evolved stars. We employ a new
calibration method for accurate circular VLBI polarimetry at
millimeter wavelengths, to present the detailed Stokes $\{I,Q,U,V\}$
properties for this feature. We analyze the fractional linear and
circular polarization as a function of projected angular distance
across the extent of the feature, and compare these measurements
against several theoretical models proposed for sharp rotations of
electric vector position angle in polarized SiO maser emission. We
find that the rotation is most likely caused by the angle $\theta$
between the line of sight and a projected magnetic field crossing the
critical Van Vleck angle for maser propagation. The fractional linear
polarization profile $m_l(\theta)$ is well-fit by standard models for
polarized maser transport, but we find less agreement for the
fractional circular polarization profile $m_c(\theta)$.
\end{abstract}


\section{Introduction}

Circumstellar SiO maser emission in commonly-observed transitions,
such as $\nu \in \{1,2\},\ J=1-0,$ or $\nu =1,\ J=2-1$, is ubiquitous
toward large-amplitude, long-period variable (LALPV) stars
\citep{hab96}. The excitation conditions for these masers place them
in the near-circumstellar environment (NCSE), between the stellar
photosphere surface and the inner dust-formation radius \citep{eli80},
but with the extent of the latter separation dependent on stellar
pulsation phase \citep{wit07}.  The maser emission is observed to be
significantly linearly polarized \citep{tro79,her06} but has a lower
level of measured circular polarization \citep{bar87,her06}. The
intrinsic high-brightness temperature and compact spatial structure of
individual SiO maser components \citep{mor79} make them valuable
scientific probes of the astrophysics of the\ NCSE. Very Long Baseline
Interferometric (VLBI) polarimetry, in particular, allows the
properties of the NCSE to be imaged at sub-mas angular resolution by
using the SiO maser components as tracers \citep{kem02}.  Particularly
important amongst the NCSE properties are the magnetic field magnitude
and distribution, both at local and global stellar scales, and its
associated dynamical influence in this environment.

Inference of magnetic field properties from the measured sub-mas SiO maser polarization however requires the solution of an inverse problem  involving the radiative transfer of SiO maser emission in full polarization. 

There remain differences in theoretical models for the transport of polarized maser emission in the limit of small Zeeman splitting, as applicable to the non-paramagnetic SiO molecule \citep{eli02,wat02}. Both collisional \citep{eli80} and radiative \citep{buj81} mechanisms have also been proposed for SiO maser pumping. To complicate matters further, the physical conditions in the NCSE include complex phenomena such as shocks, anisotropic stellar illumination of maser cells, and likely both velocity and magnetic field gradients, all of which affect the interpretation of maser polarization observations.

Spectral-line observations in full Stokes $\{I,Q,U,V\}$ with sufficient angular resolution to resolve component-level SiO polarization properties can however provide unique constraints on the theoretical and observational uncertainties enumerated above. In this paper, we report such measurements for an individual SiO maser feature displaying a distinctive observational signature, namely a rotation of approximately $\frac{\pi}{2}$ in polarization electric vector position angle (EVPA) across the component at the inner shell boundary.

The presence of such components was noted in the first VLBI polarimetry of SiO masers toward TX Cam \citep{kem97}
but has also been noted in some subsequent observations of other circumstellar SiO maser sources \citep{des00,cot06}.

Several distinct theoretical explanations have been proposed for such features.
\citet{ase05} performed an analysis of the impact of radiative anisotropy on the excitation and linear polarization of unsaturated circumstellar SiO masers. In the non-magnetic case, linear polarization tangential to the projected shell is predicted, as described also in earlier work by \citet{wes83}. In evaluating possible causes of large EVPA rotations from this default tangential orientation, \citet{ase05} considered   the impact of the Hanle effect in the presence of  magnetic fields of order 10-100 mG. They find that although the Hanle effect can produce substantial EVPA rotations for a restricted range of magnetic field geometries, the required conditions produce a predicted suppression of the masing effect; it is therefore ruled out in their analysis. The authors conclude instead that the EVPA rotation is most likely caused by a change in the radiation anisotropy conditions across the maser component, from radial anisotropy to tangential anisotropy moving outward from the inner shell radius. In this model, the magnetic field is not dynamically significant.

A second theoretical explanation involves the transport of linearly polarized maser emission in the presence of a magnetic field inclined at an angle to the line of sight. Foundational work on polarized maser radiation transfer was undertaken by \citet{gol73}. In general, the applicability of polarized maser transport solutions and their derivation is characterized by the relative magnitude of the maser stimulated emission rate $R$, radiative decay rate $\Gamma$,  Zeeman rate $g\Omega$, and bandwidth of the maser amplified radiation $\triangle \omega$ \citep{gol73}. For the case of saturated linear masers isotropically pumped over quantum number $m,$  in the asymptotic limit $\triangle \omega \gg g\Omega \gg R \gg \Gamma$, these authors derived a solution in which the measured EVPA will rotate by    $\frac{\pi}{2}$  as the angle $\theta$ between the magnetic field direction and the line of sight crosses the critical Van\ Vleck angle $\theta_F\backsimeq55\degr$, defined as $\sin^2 \theta_F=\frac{2}{3}$  \citep{gol73}. This change in $\theta$ can therefore be gradual but still result in an EVPA rotation of $\frac{\pi}{2}$. For $\theta<\theta_F$ the measured polarization EVPA is parallel to the projected magnetic field direction, and perpendicular for $\theta > \theta_F$ \citep{gol73}. \citet{eli02} noted that a transition across $\theta_F$ could be responsible for the reported $\frac{\pi}{2}$ rotations in EVPA across SiO maser features. We note that such an EVPA rotation has been detected in the polarized water maser emission toward W43A \citep{vle06}.

A third theoretical explanation for significant EVPA rotation
across SiO maser components has been presented by \citet{sok02}, as part of a larger analysis of the likely dynamical influence of magnetic fields in the evolution of late-type, evolved stars \citep{sok02a}. These authors do not support the model of a globally-organized magnetic field that is active during AGB evolution and acts dynamically to shape later planetary nebula geometry. However their work does support cool spots on AGB stellar surfaces as sites of enhanced dust formation \citep{sok99} with possible local magnetic fields of order 1-10 G \citep{sok02}. The predicted local magnetic field geometry is radial above the cool spot, but tangential closer to the photosphere \citep{sok02}. The NCSE is traversed by shocks caused by the pulsation of the central LALPV star \citep{bow88}. The contrast in magnetic field direction is expected to be enhanced by post-shock compression  \citep{sok02}. The intrinsic change in magnetic field orientation from tangential to radial over a short projected angular distance near a cool photosphere spot is accordingly proposed by \citet{sok02} as the cause of EVPA changes of $\frac{\pi}{2}$ across individual SiO maser components.

In this paper, we utilize recent algorithmic developments that allow accurate circular polarization measurement in millimeter-wavelength spectral VLBI observations \citep{kem11} to analyze the linear and circular polarimetric profile across an individual SiO component exhibiting an EVPA rotation of approximately $\frac{\pi}{2}$. We measure the fractional linear and circular polarization profiles across the feature, and assess them against the theoretical models discussed above. We find that the EVPA rotation is most likely caused by the angle $\theta$ between the line of sight and a magnetic field crossing the
critical
angle $\theta_F$ noted above.

The paper is organized as indicated below. The observations are described in Section 2 and their analysis and resulting science products presented in Section 3. The science results are discussed in Section 4, and conclusions presented in Section 5. \

\section{Observations}

The current data are from a single epoch of a more extensive   monitoring campaign using the Very Long Baseline Array (VLBA), operated by the NRAO\footnote{The National Radio Astronomy Observatory is a facility of the National Science Foundation operated under cooperative agreement by Associated Universities, Inc.}, to image the $\nu=1,\ J=1-0$ circumstellar SiO maser emission  toward the Mira variable, \object{TX Cam}. These observations were conducted at sub-milliarcsecond angular resolution and in full polarization Stokes $\{I,Q,U,V\},$ over several pulsation periods of the central star. The larger survey has been  published in total intensity by \citet{dia03} and \citet{gon10}, and in linear polarization by \citet{kem09}. 

The current epoch was observed under VLBA project code BD46AQ. The observations were scheduled on 6 February 1999 (MJD 51215),  from 0$^h$ UT to 8$^h$ UT using all ten VLBA antennas plus an  additional single VLA antenna. A total of 6.5 hours were assigned to the 43 GHz  observations, divided into on-source scans of 13-minute duration, and comprising the following  total integration times: i)  seventeen  scans
on the target source, TX Cam; ii) seven scans on the continuum extragalactic calibrator J0359+509; and iii) one scan each on the continuum extragalactic calibrators 3C454.3 and J0609-157. The scans on TX\ Cam and J0359+509 were distributed as evenly as possible to maximize $uv-$coverage. Approximately 86.7\% of the elapsed schedule time was on-source, the remainder was allocated for antenna slewing, system initialization, or to balance the tape resources allocated at that time.

The $v=1,J=1-0$ SiO maser transition was observed in a 4 MHz baseband, and centered on a systemic LSR velocity of +9 kms$^{-1}$  for TX\ Cam. The adopted rest frequency was 43.122027 GHz.
No real-time Doppler tracking was employed about the mean Doppler shift computed for the mid-point of the schedule and the array; these corrections were applied in post-processing, consistent with standard spectral-line VLBI practice \citep{dia89}. The data were sampled in 1-bit quantization, and correlated in full polarization over a maximum possible 128 frequency channels available at that time, yielding a nominal channel width of 31.25 kHz. The correlator accumulation interval was 4.98 s.     

\section{Results}

The data were reduced using the method described by \citet{kem11}
for accurate circular VLBI polarimetry at millimeter wavelengths. The primary science products resulting from the reduction are image cubes in each of Stokes $\{I,Q,U,V\}$, at a pixel spacing of 50 $\mu$as (2048 x 2048) on the image tangent plane, with one image per sampled frequency channel over the inner 113 frequency channels in the spectrum. A common restoring beam of size $540 \times 420\ \mu$as
at a position angle of 20\degr\ was adopted across all epochs of the larger survey, as described by \citet{kem09}. The absolute electric-vector position angle (EVPA) of the linearly-polarized emission was established from associated VLA observations, relative to the primary polarization EVPA calibrator, 3C138, as described by \citet{kem09}.
The residual error in the absolute EVPA determination is estimated  to be $\sim10\degr-20\degr$ peak-to-peak \citep{kem09}.

\begin{figure}
\includegraphics[angle=-90,scale=0.40]{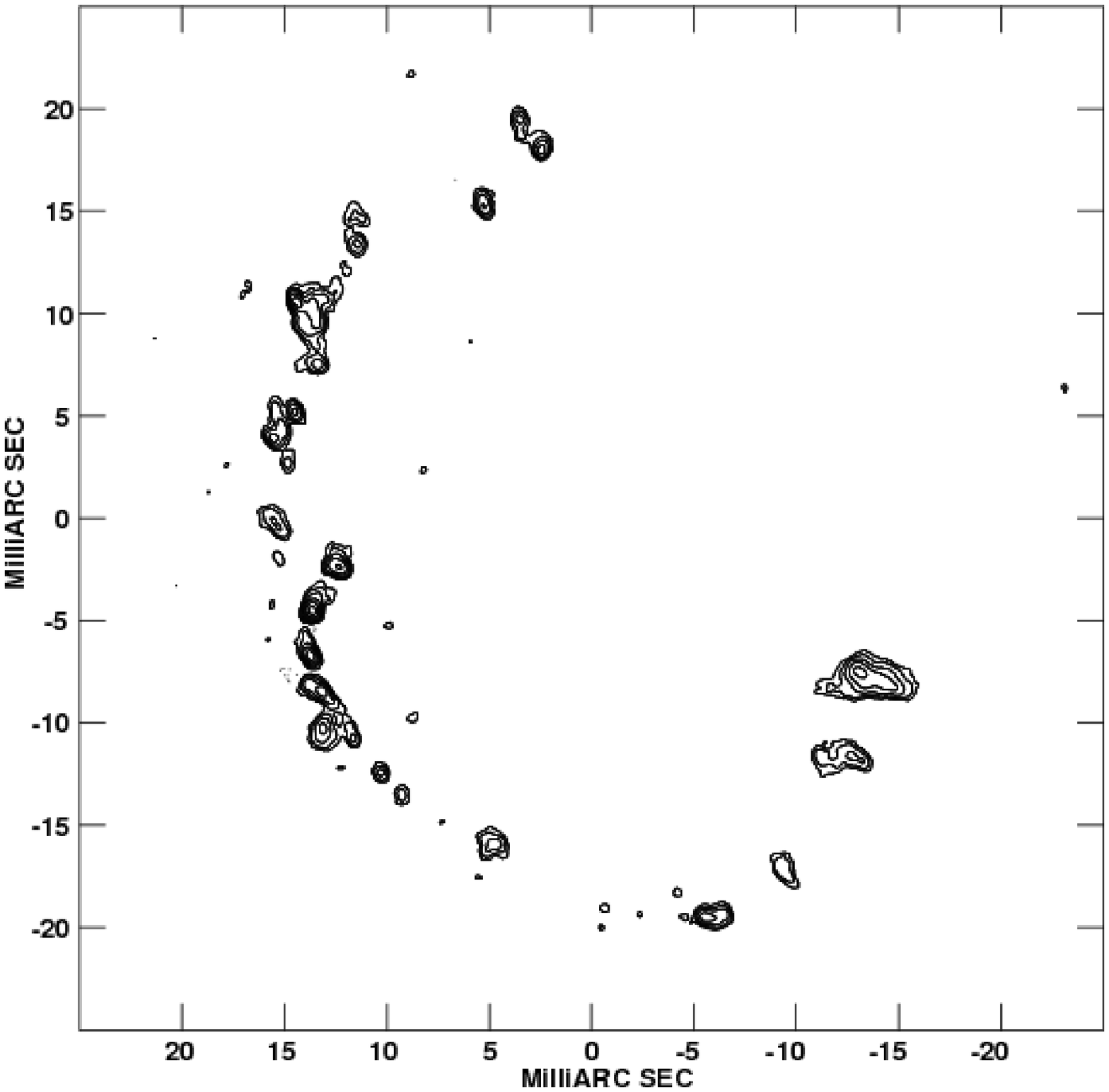}
\caption{Stokes $I$ contour image of the $\nu=1,\ J=1-0$ SiO maser emission toward TX Cam, plotted as the zeroth moment over frequency of the image cube. In these averaged units, the contour levels are at $\{-10,\ -5,\ 5,\ 10,\ 20,\ 40,\ 80,\, 160,\ 320\} \times \sigma$, where $\sigma$ is the off-source rms of 2.1982 mJy/beam. The angular coordinates are in mas from the center of the sub-image enclosing the projected SiO maser ring.}
\label{fig-sqash-i}
\end{figure}

The zeroth moment, over frequency, of the full Stokes $I$ image cube is shown in Figure~\ref{fig-sqash-i}. The counterpart Stokes $V$ image is shown in Figure~2\notetoeditor{reference fig-sqash-v does not produce correct numbering here}. The linearly-polarized intensity $P=\sqrt{(Q^2+U^2)} $,  derived directly from the zeroth-moment images in Stokes $Q$ and $U$, is depicted in Figure~3\notetoeditor{reference fig-sqash-p does not produce correct numbering here.}. In Figure~4\notetoeditor{reference fig-sqash-pchi does not produce correct numbering here}, the  Stokes $I$ image is shown  overlaid with vectors proportional in length to $P$ and drawn at a position angle equal to the absolute EVPA of the linearly-polarized emission.

This paper concerns a feature located near relative coordinates (-15, -7.5) mas in Figure~4\notetoeditor{reference fig-sqash-pchi does not produce correct numbering here} that has an approximate $ \frac{\pi}{2}$ rotation in EVPA
  across the feature. We note that the image cubes lack absolute astrometric coordinates due to the use of VLBI phase self-calibration. This component, near the southwest circumstellar boundary, is shown in Stokes $I$ and $P$ in Figure~\ref{fig-sqrev-pchi}, averaged over frequency as in earlier Figures. As noted earlier, it is not uncommon to find adjacent individual SiO components with large (or perpendicular) rotations in relative EVPA. In the current epoch, we choose this particular individual feature for further analysis because it spans a relatively large range in velocity and has good SNR.

The individual frequency channel images across this feature, labeled by line-of-sight velocity (in km/s), are plotted in Stokes $I$ and $P$ in Figure~\ref{fig-reversal-pchi}, and in Stokes $V$ in Figure~\ref{fig-reversal-v}. 

In the absence of knowledge of the astrometric position of the central star relative to the SiO emission, we adopt the approximation that radially-extended maser features  point back to the photosphere, an assumption for SiO masers discussed by \citet{zha11}. The corresponding vector obtained by fitting a straight line to the projected coordinates of the peak Stokes $I$ component brightness across the velocity extent of the $\frac{\pi}{2}$ EVPA rotation feature is drawn  as an arrow at upper left in Figures~\ref{fig-sqrev-pchi},~\ref{fig-reversal-pchi}, and \ref{fig-reversal-v}, and annotated accordingly as the approximate assumed direction toward the photosphere.

The integrated mean-intensity spectrum, computed across the full image region in Figure~\ref{fig-sqrev-pchi} enclosing the  $\frac{\pi}{2}$ rotation feature, is plotted in Stokes $I$, Stokes $V$, and linearly-polarized intensity $P=\sqrt{Q^2+U^2}$ in Figure~\ref{fig-spectrum-ivp}.

\section{Discussion}

The large-scale morphological properties of the $\nu=1,\ J=1-0$ SiO maser emission distribution at this epoch in Stokes $I$ and $P$  are consistent with the summary results for the broader monitoring campaign, as discussed by \citet{dia03} and \citet{kem09}.
The total intensity distribution, depicted as an average over frequency in Figure~\ref{fig-sqash-i}, shows the projected ring-like shell morphology frequently found for circumstellar SiO maser emission \citep{dia94}. The corresponding linearly-polarized intensity, shown in Figures~3 and 4\notetoeditor{references fig-sqash-p and fig-sqash-pchi do not produce correct numbering here}, has the characteristic tangential distribution of EVPA found to be persistent across the broader monitoring campaign \citep{kem09}.
However\ this tangential polarization morphology is not universal for circumstellar SiO masers as a class
\citep{cot09}.

The individual component studied in this paper is located at the projected shell boundary in the southwest region of the overall SiO maser distribution. As shown  in Figure~\ref{fig-sqrev-pchi}, there is a rotation of approximately $\frac{\pi}{2}$ in EVPA across the component, with a tangential orientation on the inner shell boundary and a radial orientation at a larger  projected angular distance from the central star (which lies  toward the northeast in Figure~\ref{fig-sqrev-pchi}).  

The component is elongated  in total intensity along the radial axis. This is consistent with tangential amplification of the underlying maser emission, as described by \citet{dia94}, and consistent with radial shock acceleration predicted in the circumstellar shells of LALPV stars \citep{hum02,gra09}. The total intensity contour images of each frequency channel across the feature (Figure~\ref{fig-reversal-pchi}) show that the projected center of the Stokes $I$ emission moves inward toward the central star with decreasing LSR velocity, as expected for a positive radial gradient in the outflow velocity  at the position of this SiO maser component in this shell.

The linear polarization channel images shown in Figure~\ref{fig-reversal-pchi} span the feature in velocity. They indicate an abrupt transition in EVPA of $\frac{\pi}{2}$ near $V_{LSR}\sim+6.2$ km/s and a corresponding local minimum in linearly-polarized intensity near this point. The corresponding circular polarization channel images in Figure~\ref{fig-reversal-v} show a sharp decline in peak Stokes $V$ with decreasing LSR velocity, i.e. in the direction toward the inner projected shell boundary. 

The fractional linear ($m_l$) and circular $(m_c)$ polarization magnitude profiles across the feature are plotted in Figure~\ref{fig-profile-zplot}, for all measurements with an SNR exceeding 3. The magnitudes of the fractional polarizations are measured at the single pixel position of  maximum Stokes $I$ in each channel image. The $x-$ordinate in this plot is the projected angular separation $d$ (mas) from the component peak in the channel image at $V_{LSR}=+7.91$ km/s in the upper left panel of Figure~\ref{fig-reversal-pchi}. In this sense, the projected angular separation increases toward the central star - with decreasing $V_{LSR}$ due to the radial velocity gradient discussed above.

The single-pixel measurements of $m_l$ include the effects of spatial
linear depolarization arising from convolution by the synthesized beam
during image formation. The magnitude of this effect can be assessed
for a given deconvolved source component size $\sigma_m$, a
synthesized geometric beamwidth $\sigma_b$, and an adopted linear rate
of change of EVPA $\alpha={\dot \chi}$ with angular spatial scale in
the image. The angular sizes $\sigma$ are expressed here as the full
width at half maximum intensity (FWHM). The components in
Figure~\ref{fig-sqrev-pchi} with the largest apparent values of
$\alpha$ are at $V_{LSR}=+6.18$ km/s and $V_{LSR}=+5.96$ km/s.  The
mean deconvolved source component size across the minor axis (which is
almost perpendicular to the EVPA), for these velocities is $\sigma_m
\sim 1.2$ mas. The geometric synthesized beamwidth for the current
data is $\sigma_b \sim 0.48$ mas (see above). The linear beam
depolarization arising in this case is approximatly $m_l' \simeq
\frac{\sin(\alpha \sigma_b)}{\alpha \sigma_b}m_l=\beta m_l$, with
$\alpha=\frac{\triangle \chi}{\sigma_m}$. Adopting ${\triangle
\chi}_2=\frac{\pi}{2}$ and ${\triangle \chi}_4=\frac{\pi}{4}$ produces
$\beta=0.94$ and $\beta=0.98$ respectively, with values closer to
unity for the other components in Figure~\ref{fig-reversal-pchi}, due
to significantly lower apparent values of $\alpha$. Thus, for the
single-pixel measurements of $m_l$ presented here, spatial beam
depolarization is not believed to substantially affect the current
analysis.

For the case of the single-pixel fractional circular polarization measurements, $m_c$, it is similarly appropriate to consider the effect of depolarization arising from averaging over frequency. As noted earlier, the nominal channel increment in the data is 31.25 kHz; in uniform spectral weighting the effective FWHM spectral resolution is $\sigma_v \sim 0.26$ km/s. The mean observed FWHM component line-width in Stokes $I$ over the feature is $\sigma_{obs} \sim 0.9$ km/s.
To first-order, for a classical Zeeman "S-curve"  the peak Stokes $V$ obeys the proportionality $V_{max} \propto \frac{\triangle_z}{\sigma}$, for a Zeeman splitting $\triangle_z$ and component line-width $\sigma$. Convolution by the instrumental frequency response therefore scales $V_{max}$ by
a factor $\eta\sim \sqrt{1-(\frac{\sigma_v}{\sigma_{obs}})^2}$.
For the values of $\sigma_v$ and $\sigma_{obs}$ for the current data, $\eta \sim 0.96$, which does not significantly affect the results presented in the current work.
This calculation is illustrative only however, as the Stokes $V$ component profiles do not take simple Zeeman form, as evident from the integrated spectrum plotted in Figure~\ref{fig-spectrum-ivp}. However, we believe the calculation is nonetheless representative of the magnitude of the circular depolarization effect. We discuss the broader question of depolarization caused by line-of-sight integration below. 

The asymptotic linear polarization solution for saturated, $m-$isotropic linear masers derived by \citet{gol73} for the case where $\triangle \omega \gg g\Omega \gg R \gg \Gamma$ predicts a fractional linear polarization dependence on the angle $\theta$ between the magnetic field and the line of sight taking the form \citep{gol73}:

\begin{eqnarray}
m_l(\theta)&=& \frac{2-3\sin^2\theta}{3\sin^2 \theta}\ {\rm for\ }\theta \ge\theta_B\nonumber\\
m_l(\theta) &=& 1\ {\rm for\ }\theta \le \theta_B
\label{eq-gkk}
\end{eqnarray}
where $\tan^2 \theta_{B} = \frac{1}{2}$. In this solution the measured EVPA rotates by $\frac{\pi}{2}$ at $\theta_F$, where $\sin^2 \theta_F=\frac{2}{3}$ \citep{gol73}. We denote this solution as GKK in what follows. We note that the GKK solution was derived specifically for a $J=1-0$ transition, as applies to the transition observed here.

We model the fractional linear polarization data across the SiO maser feature by fitting the lowest-order polynomial form of $\theta(d)$ on the projected angular distance $d$ across the feature that yields a reasonable fit.     For the current feature this is a quadratic fit:

\begin{equation}
\theta(d)=a(d^2-d_f^2)+b(d-d_f)+\theta_F
\end{equation}

where $d_f=2.822$ mas is the projected angular separation at which the EVPA rotates by $\frac{\pi}{2}$ in the measured data - here chosen to be the channel at $V_{LSR}=6.18$ km/s. The free parameters in the fit are $a$ and $b$.
In Figure~\ref{fig-profile-zplot} we plot the best chi-square fit of the measured $m_l$ data to the GKK solution as a dashed line.
The
associated solution for $\theta(d)$ is plotted in Figure~\ref{fig-profile-chiplot}.

The fit to $m_l$ shows broad agreement with the functional form of the GKK solution. In this model there is a gradual change in the angle $\theta$ between the line of sight and the projected magnetic field over the range 37.5\degr\  to 70\degr\ across the feature (with increasing $d$).
The angle $\theta$ crosses the Van Vleck angle $\theta_F$ at $d=d_f$ near $V_{LSR}=+6.2$ km/s.
We note that for this feature the lower bound of 37.5\degr\ is consistent with the lack of a stable solution for $\theta < \theta_B$ predicted by \citet{eli96}.

If we assume that $m_l(\theta)$ originates from a GKK solution then we can derive the dependence of fractional circular polarization on $\theta$ by direct inversion of Equation~\ref{eq-gkk} as:

\begin{equation}
\cos \theta = \sqrt{1 - \frac{2}{3(m_l+1)}}
\label{eq-gkk-inv}
\end{equation}

We plot the resulting dependence of measured fractional circular polarization on $\theta$ in the form $m_c(\cos \theta)$ in Figure~\ref{fig-profile-mcplot}. Here, the measured values of $m_l$ are used to compute $\cos \theta$ using Equation~\ref{eq-gkk-inv}; the matching values of $m_c$ are then plotted against the derived  $\cos \theta$ abscissa. The plot shows an  increase of $m_c$ with $\cos \theta $ over a relatively narrow range of angles $\theta$ for which the projected magnetic field is closest to the line of sight.

We can assess the measured component-level polarization properties presented here relative to the different theoretical models that have been proposed to explain EVPA rotations of approximately $\frac{\pi}{2 }$ across individual SiO maser features, described in the Introduction.
However, there are important caveats that apply; we know this region has magnetic field and velocity gradients and is in a shock-traversal region; many theories were developed within more idealized conditions, by necessity.
Furthermore, our observations are integrated over the synthesized angular beamwidth along lines of sight
within this complex environment . We do not believe this affects substantially the primary conclusions of this paper however, but more complex three-dimensional modeling of this region is planned in future work.

The non-magnetic ($g\Omega=0$) model of \citet{ase05}
explains the EVPA rotation as due to changes in radiative isotropy across the SiO maser feature. We believe that the current data do not provide support for this model. If the tangential linear polarization at the inner edge of the shell is predominantly caused by $m$-anisotropic pumping in a non-magnetic environment, then the measured circular polarization would have to arise from non-Zeeman effects \citep{wat09}; in this case the inter-conversion of linear polarization to circular polarization is due to a change in optical axes along the propagation path. In a non-magnetic model, we might expect from first principles that non-Zeeman circular polarization would be found preferentially at positions with the greatest  EVPA rotation rate; however the circular polarization is not pronounced at this position in our data. We also note that the greatest fractional linear polarization is measured in our data at the feature position that is furthest from the central star. Further, the relatively close agreement with the GKK functional form for $m_l(\theta)$ argues against a model of EVPA rotation due primarily to changes in radiation isotropy. For these reasons we believe that our data do not provide support for the EVPA rotation model described by \citet{ase05}.

In considering models with non-zero magnetic fields $g\Omega \ne 0$, we believe that contemporary observational evidence suggests that circumstellar SiO masers are in the regime $g\Omega \gg R \gg \Gamma$ or $g\Omega > R \gg \Gamma$ \citep{kem09}. Evidence in support of the partial subsidiary condition $g\Omega \gg R$ is provided by \citet{wat09}.
  This condition excludes intensity-dependent non-Zeeman circular polarization that is possible if $g\Omega \sim R$ \citep{ned94}. The condition $g\Omega \gg R$ also ensures that the magnetic field is always either parallel or perpendicular to the measured EVPA (as in the GKK model presented above), but not necessarily at the same value of the Van\ Vleck angle $\theta_F$ \citep{wat09} or with the same functional form for $m_l(\theta)$.

Other sources of non-Zeeman circular polarization are possible for maser models with $g\Omega \ne 0$, due also to differences between the direction of maser linear polarization and the optical axes along the propagation path, in this case caused by changes in magnetic field direction or Faraday rotation \citep{wie98,wat09}.
It is important to consider whether these sources of non-Zeeman circular polarization could explain our current observations. For this mechanism, we expect a correlation between fractional circular polarization and fractional linear polarization, and that the greatest circular polarization will occur at positions with the greatest rate of change of magnetic field direction \citep{wat09}. Our current data are inconclusive on this point. There is a partial  correlation between $m_c$ and $m_l$, and within the constraint $m_c \le \frac{m_l^2}{4}$ put forward by \citet{wie98} for this mechanism, but over too small a number of samples to allow a robust statistical conclusion from these data (see Figure~\ref{fig-profile-zplot} for reference). In contrast, the magnetic field angle gradient plotted in Figure~\ref{fig-profile-chiplot} (derived from the GKK fit) is anti-correlated with $m_c$. 
 
The linear polarization profile shown in Figure~\ref{fig-profile-zplot} can also be examined in terms of the information it provides on predicted maser saturation in different models of maser polarization propagation. The GKK solutions assume strong saturation $R \gg \Gamma$.  In the work of \citet{eli96} a linear polarization solution of GKK form can be attained well before saturation; in contrast the models of \citet{wat01} require strong saturation in a uni-directional $J=1-0$ maser to achieve the value of $m_l \sim 0.7$ reported here.  However, we do note that the latter model assumes $m-$isotropic pumping. In light of these aggregate predictions, we believe our linear polarization profile is consistent with these masers being saturated.

This admits an interpretation in which our linear polarization data are  explained as resulting from saturated maser emission $R \gg \Gamma$ from a region threaded by a magnetic field  that changes direction smoothly relative to the line of sight across the maser feature. The angle between the magnetic field and line of sight is approximately 37.5\degr \ at the furthest point from the star, and approximately 70\degr\ at the inner shell boundary of the feature. The relative orientation crosses the critical Van Vleck angle $\theta_F$ within this region, causing an abrupt $\frac{\pi}{2}$ change in measured\ EVPA \citep{eli02}. We know from earlier work \citep{kem09,cot08} that individual maser motions appear to be influenced by individual magnetic field lines. There are also very plausible mechanisms for producing such field curvature, even if only local fields are considered such as those associated with proposed AGB cool spots, as noted earlier \citep{sok02}.

Our measured circular polarization profile raises several theoretical issues. The GKK solution used earlier makes no prediction about circular polarization - it is identically zero at line center in their asymptotic solutions \citep{wat09}.
The foundational work of GKK has been generalized to a wider range of parameter space in subsequent studies however \citep{eli96,wat01}, and we examine those predictions here.

The current data show a possible increase in $m_c$ with $\cos \theta$, perhaps of linear form, but with a sharper than expected fall-off with increasing $\theta$. 
These conclusions are tempered by the limited number of points however, and clearly further data are needed. A $\cos \theta$ functional dependence occurs for thermal Zeeman emission, and for the case of widely-separated Zeeman maser components \citep{gol73}. The work of \citet{eli96} predicts a $m_c \propto \cos^{-1} \theta$ dependence for the small Zeeman-splitting case. \citet{wat01} predict a circular polarization profile over $\theta$ that changes in shape as a function of degree of saturation $\frac{R}{\Gamma}$. A linear relation between $m_c$ and $\cos \theta$ is not predicted in this theory except for highly unsaturated emission; for increasing saturation $m_c$ increases sharply toward a peak in the region $\cos \theta \le 0.5$, then declines to zero as expected at $\theta=\frac{\pi}{2}$ \citep{wat01}. However, as noted earlier, this model is for $m-$isotropic pumping.
Further observations are needed to elucidate the nature of the circular polarization profile.

\section{Conclusions}

We have examined an individual SiO maser feature that shows an EVPA rotation of approximately $\frac{\pi}{2}$ over the projected extent of the emission. From our analysis, we find:

\begin{enumerate}
\item{The fractional linear polarization, as a function of the angle $\theta$ between the line of sight and the magnetic field, reproduces the functional form of the maser polarization radiative transfer solution derived by \citet{gol73} for the case $\triangle \omega \gg g\Omega \gg R \gg \Gamma.$ A simple model is adopted for the relation between $\theta$ and the projected angular separation across the feature.}

\item{The rotation in EVPA by approximately $\frac{\pi}{2}$ is explained within this model by $\theta$ crossing the Van Vleck angle $\theta_F$ near the mid-point of the component.}

\item{Further observations are needed to clarify the functional dependence of circular polarization on $\cos \theta$.}

\end{enumerate}


\acknowledgments

We are grateful to our colleagues for their comments on earlier drafts of this paper. We  also thank the journal referee for comments that improved the clarity and scientific content of the manuscript.

{\it Facilities:} \facility{VLBA}.


\clearpage

\begin{figure}
\label{fig-sqash-v}
\includegraphics[angle=-90,scale=0.40]{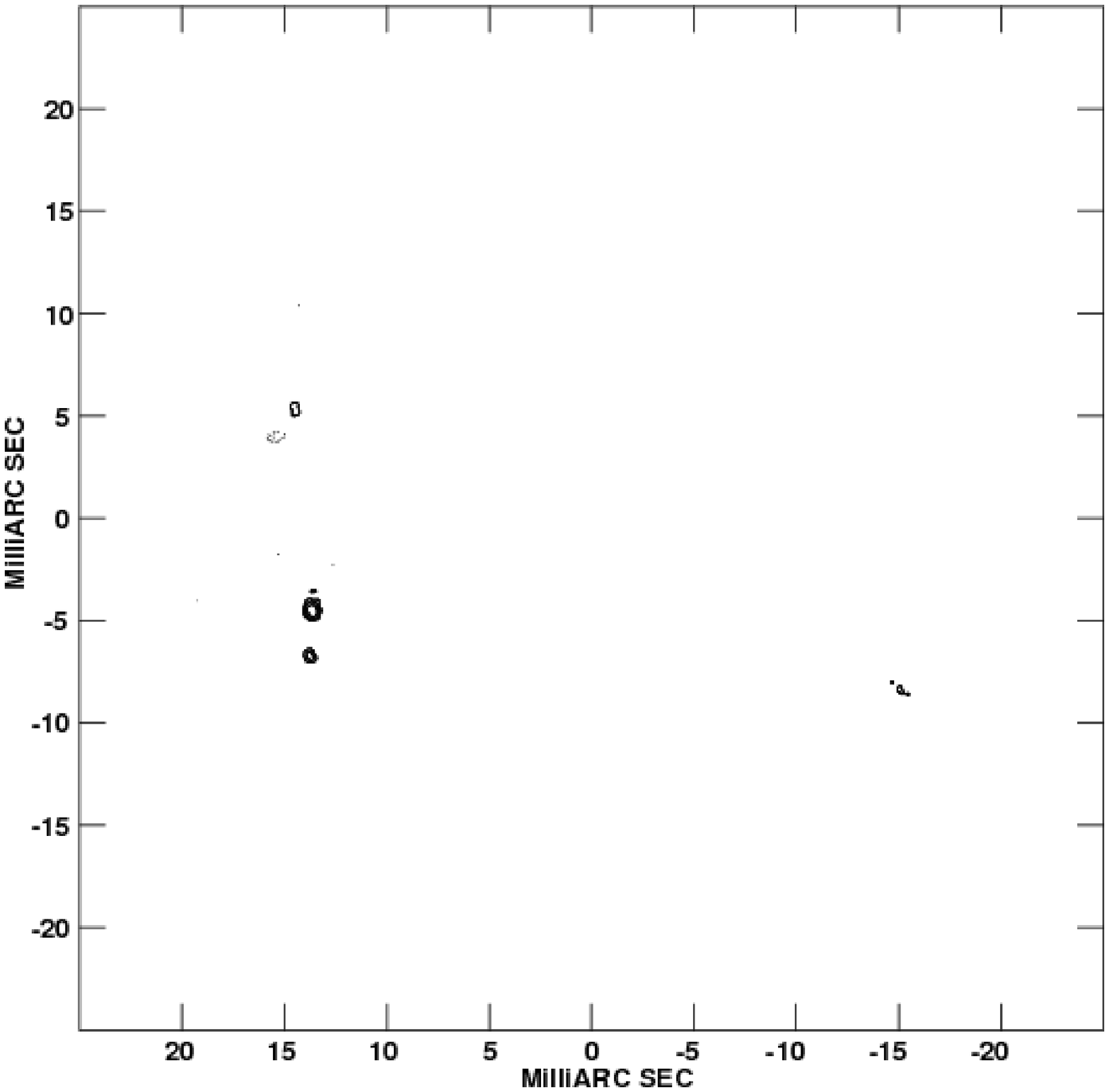}
\caption{Stokes $V$ contour image of the $\nu=1,\ J=1-0$ SiO maser emission toward TX Cam, plotted as the zeroth moment over frequency of the image cube. In these averaged units, the contour levels are at $\{-160,\ -80,\ -40,\ -20,\ -10,\ -5,\ 5,\ 10,\ 20,\ 40,\ 80,\, 160\} \times \sigma$, where $\sigma$ is the off-source rms of 1.7113 mJy/beam. The angular coordinates are in mas from the center of the sub-image enclosing the projected SiO maser ring, aligned with Figure~\ref{fig-sqash-i}.}
\end{figure}


\begin{figure}
\label{fig-sqash-p}
\includegraphics[angle=-90,scale=0.40]{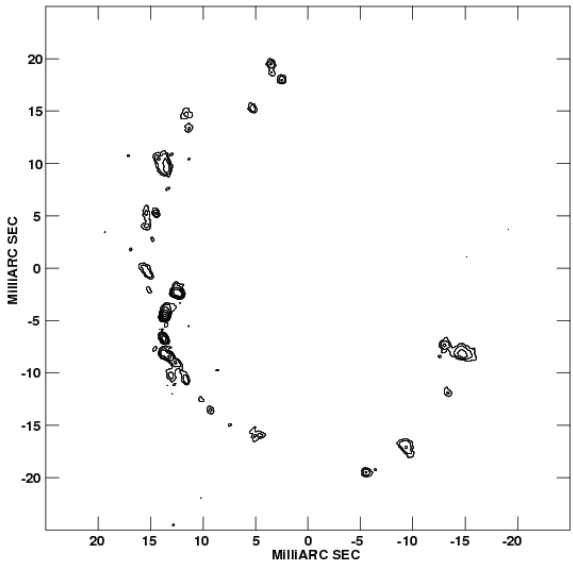}
\caption{Stokes $P$ contour image of the $\nu=1,\ J=1-0$ SiO maser emission toward TX Cam, plotted as the zeroth moment over frequency of the image cube. In these averaged units, the contour levels are at $\{7.5,\ 15,\ 30,\ 60,\ 120,\, 240,\ 480\} \times \sigma$, where $\sigma$ is the off-source rms of 1.1954 mJy/beam. The angular coordinates are in mas from the center of the sub-image enclosing the projected SiO maser ring, aligned with Figure~\ref{fig-sqash-i}.}
\end{figure}

\begin{figure}
\label{fig-sqash-pchi}
\includegraphics[angle=-90,scale=0.40]{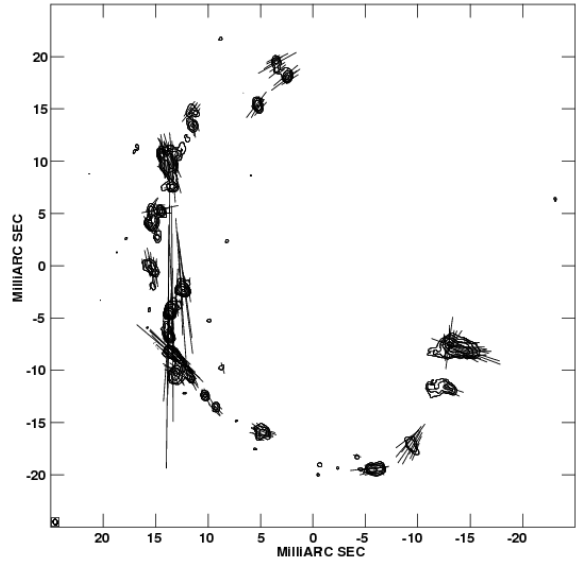}
\caption{Stokes $I$ zeroth-moment image over frequency, plotted at the contour levels in Figure~\ref{fig-sqash-i}; the overlaid vectors are drawn with position angle equal to the absolute EVPA of the underlying zeroth-moment linearly-polarized intensity $P$ and with length proportional to $P$ such that $P$= 16 mJy/beam has length 1 mas.}
\end{figure}

\begin{figure}
\includegraphics[angle=-90,scale=0.35]{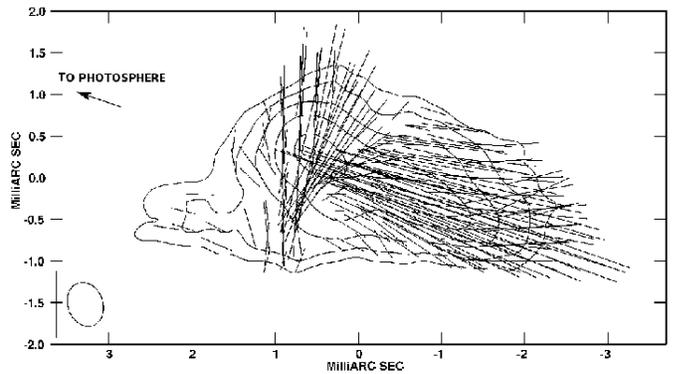}
\caption{A sub-image from Figure~4 enclosing the
south-west component on the projected shell boundary with a
$\frac{\pi}{2}$ rotation in EVPA. The Stokes $I$ contour levels are as
used in Figure~4, but the linearly-polarized
zeroth-moment intensity $P$ is drawn with vector length on a scale
where $P$ = 26.7 mJy/beam has length 1 mas.}
\label{fig-sqrev-pchi}
\end{figure}

\begin{figure*}
\centering
\begin{tabular}{cc}
\includegraphics[angle=0,scale=0.25]{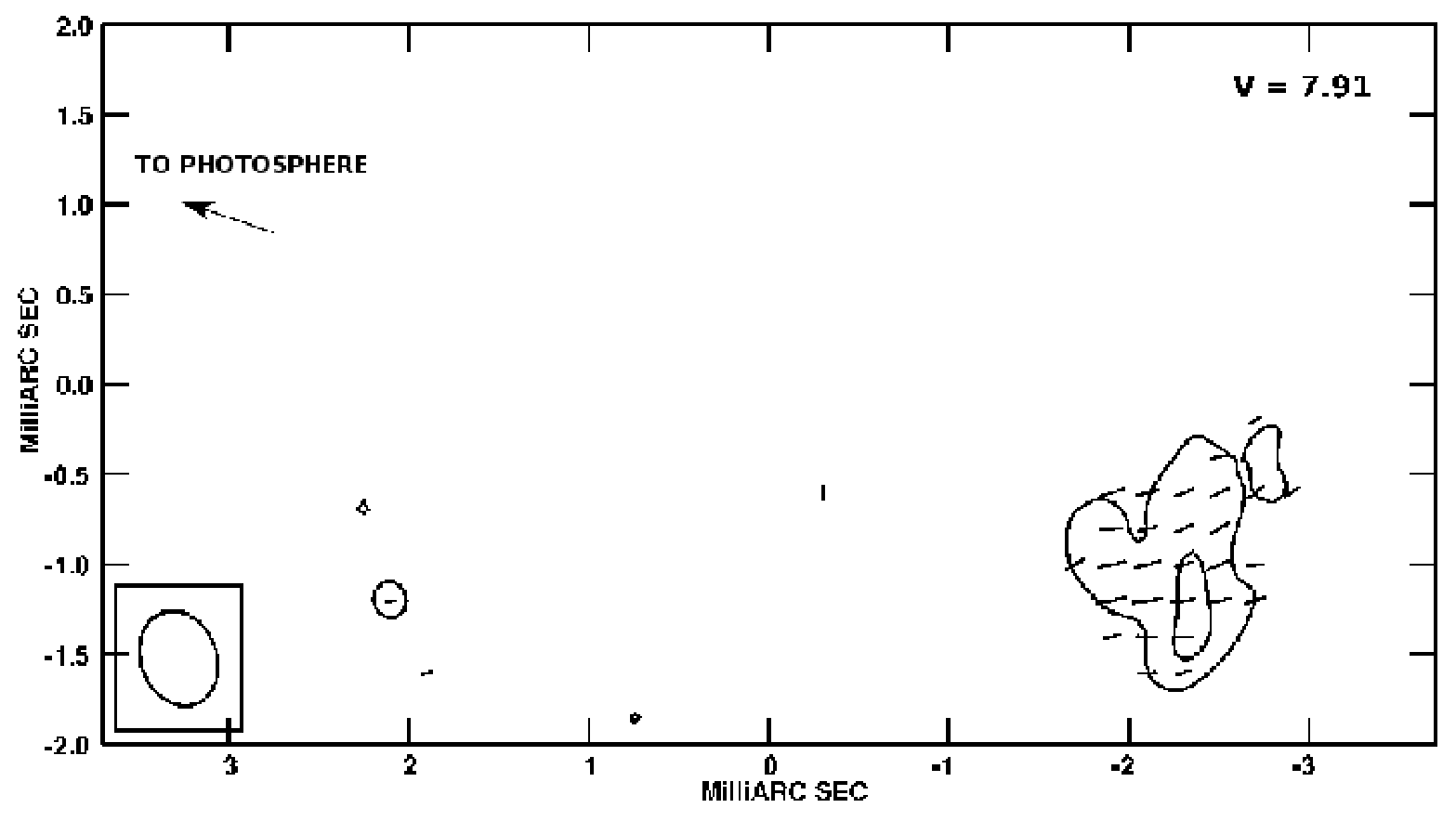} &
\includegraphics[angle=0,scale=0.25]{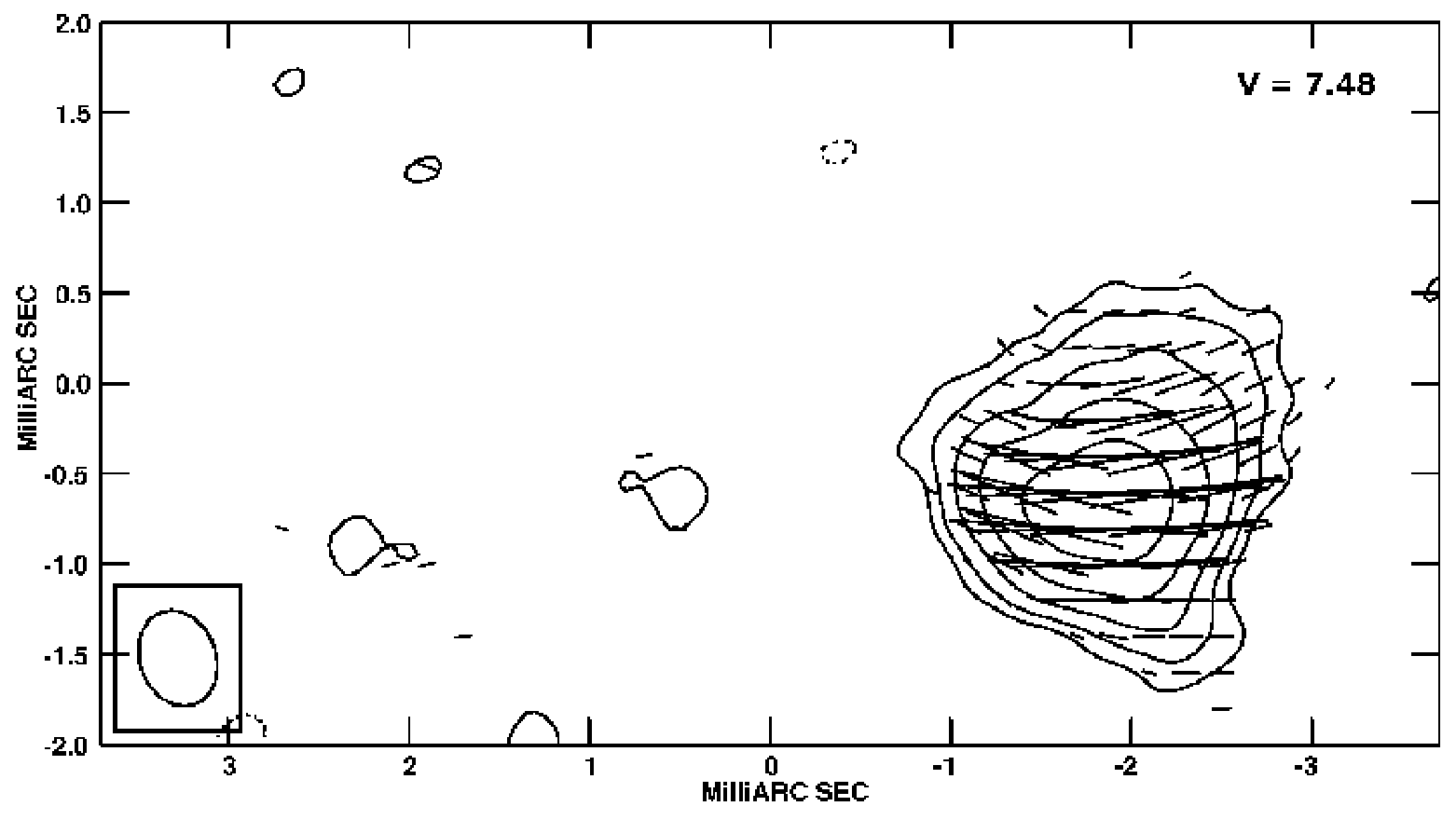} \\
\includegraphics[angle=0,scale=0.25]{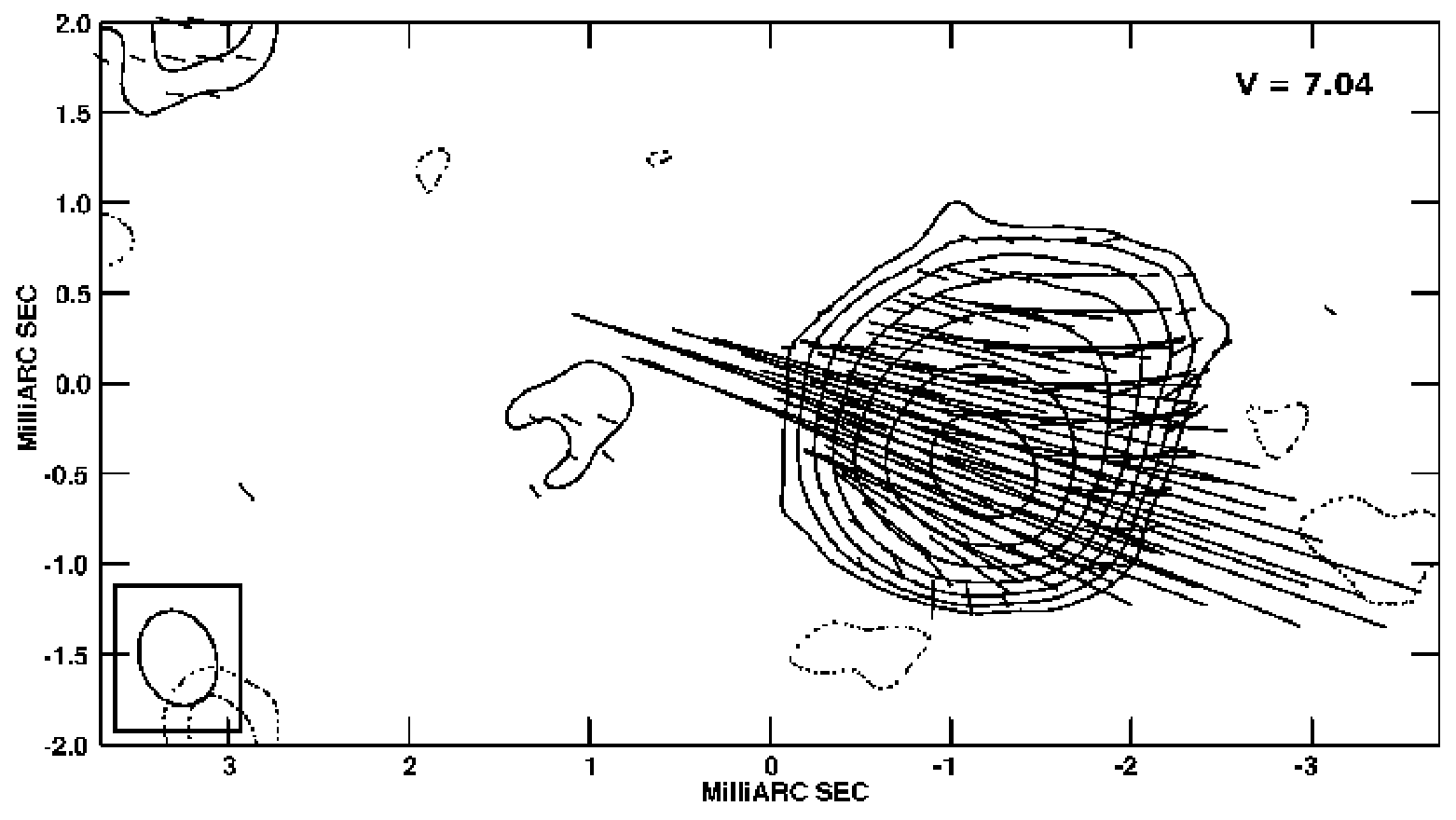} &
\includegraphics[angle=0,scale=0.25]{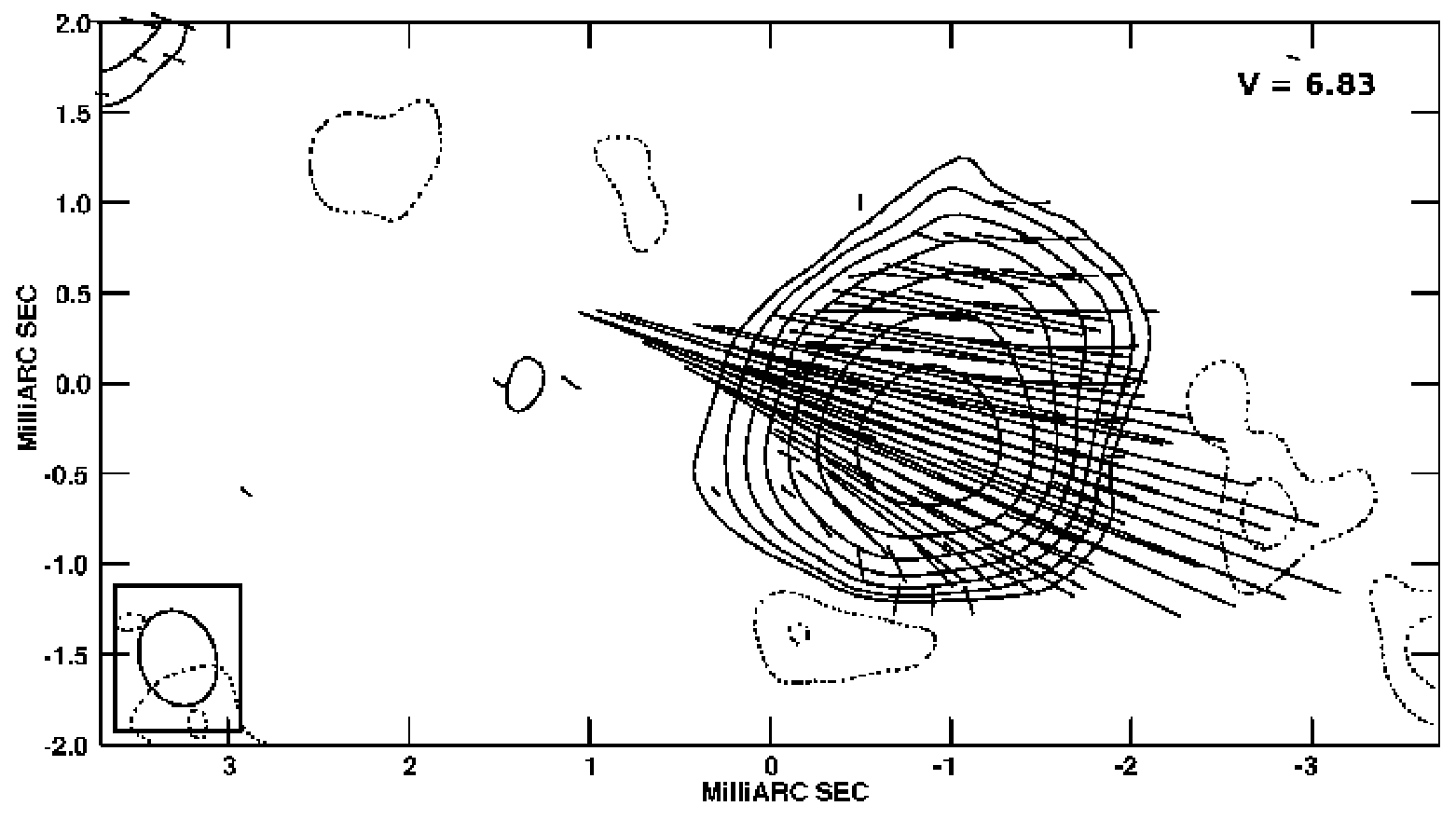} \\
\includegraphics[angle=0,scale=0.25]{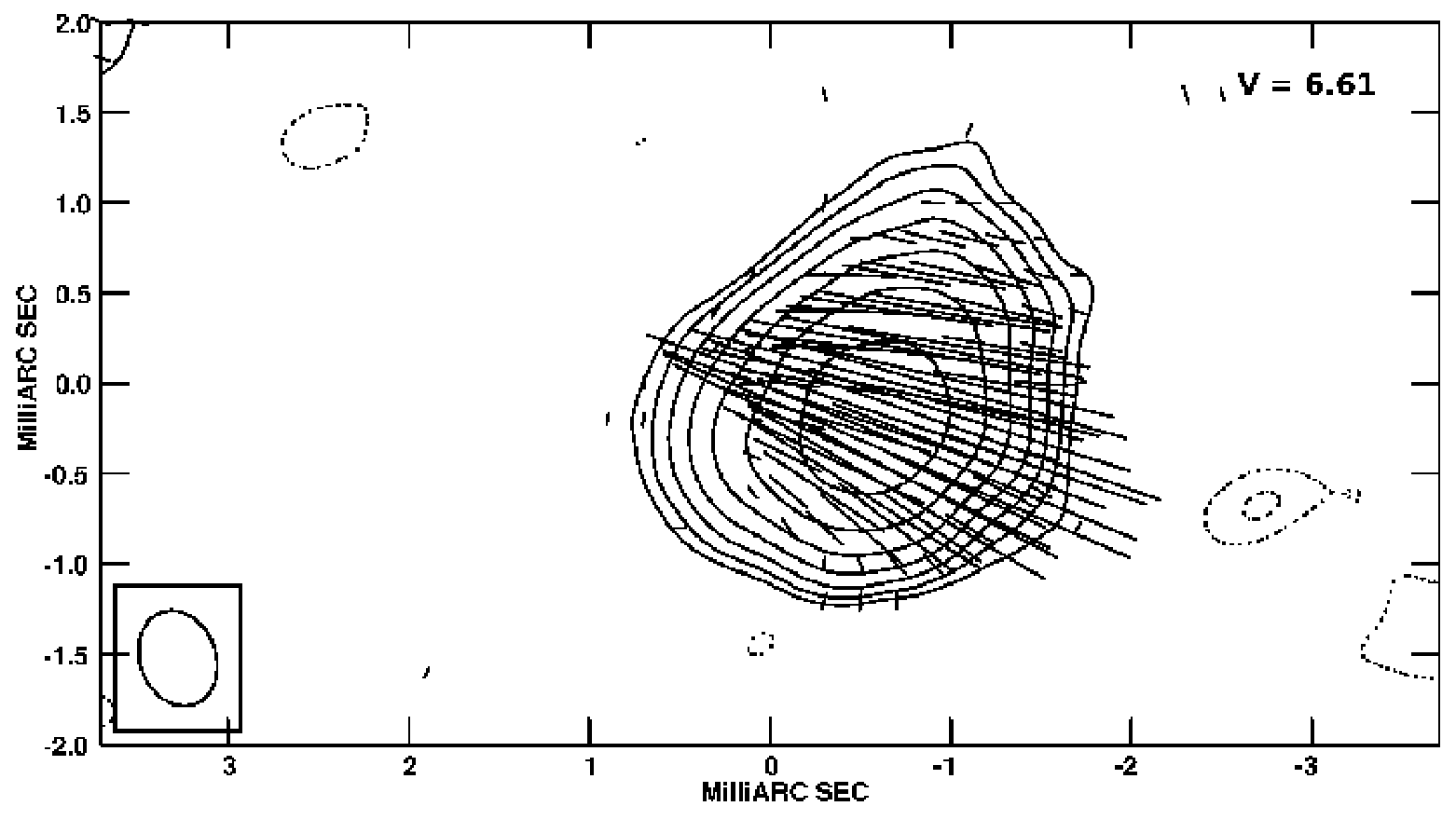} &
\includegraphics[angle=0,scale=0.25]{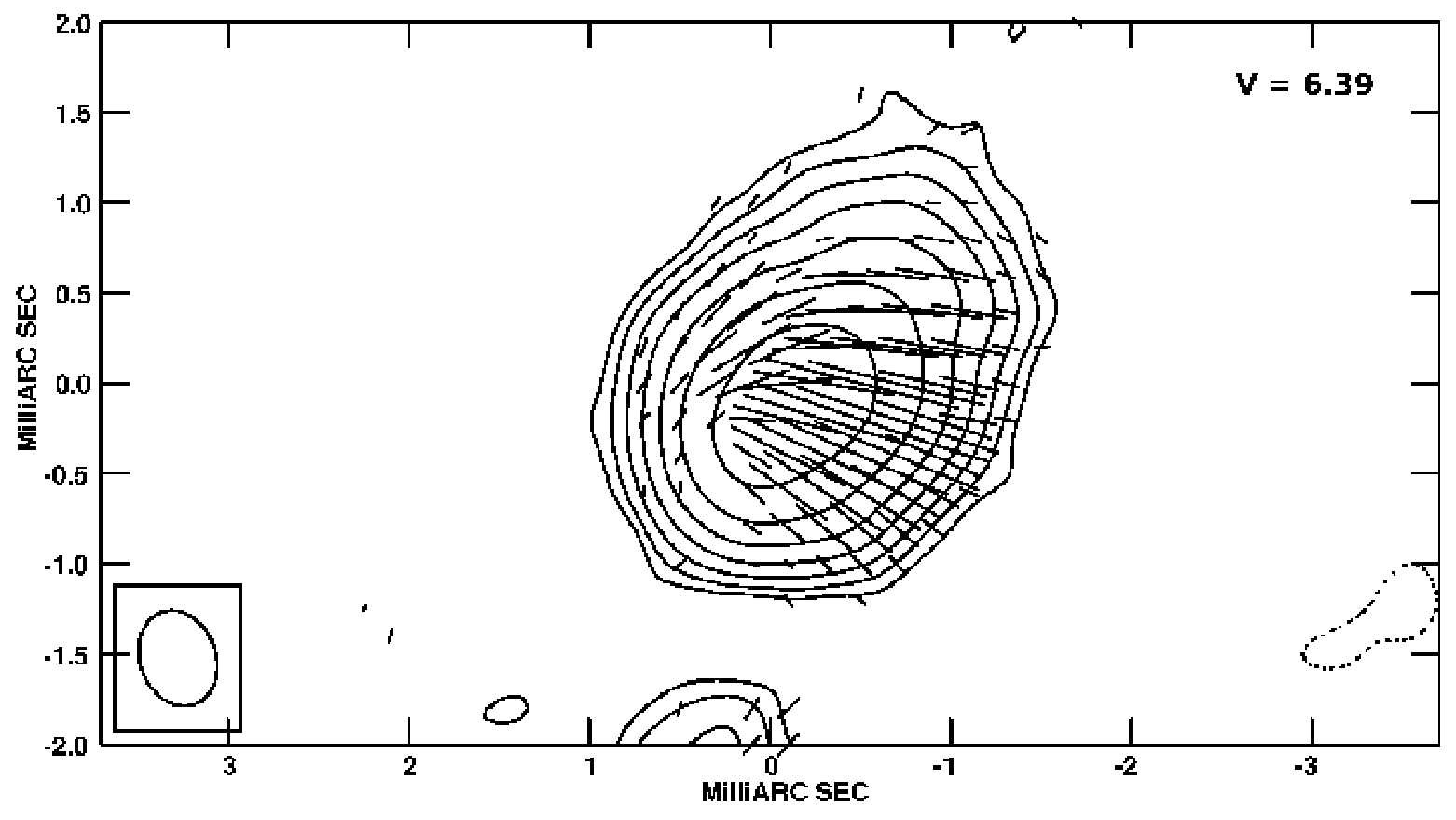} \\
\includegraphics[angle=0,scale=0.25]{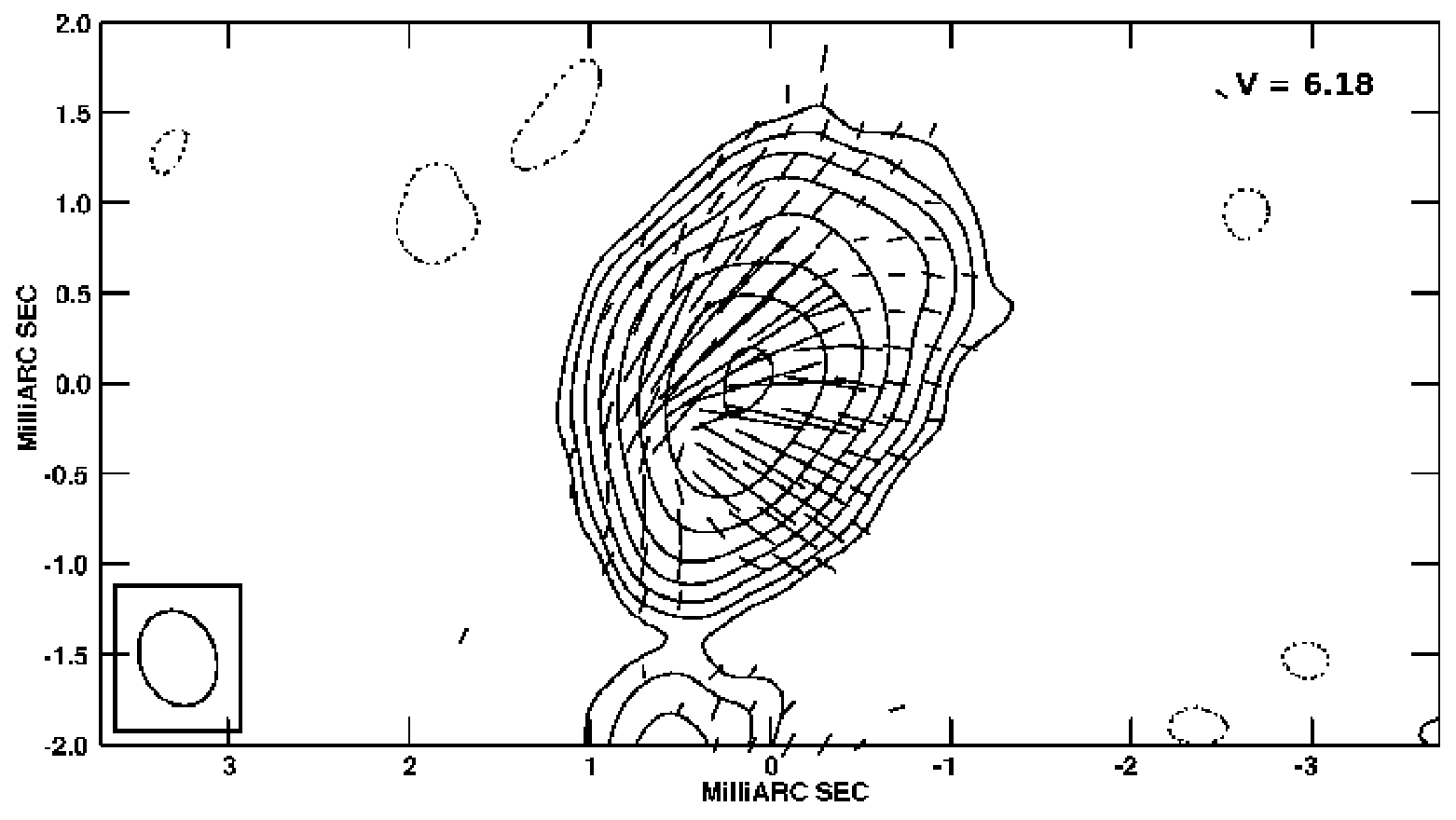} &
\includegraphics[angle=0,scale=0.25]{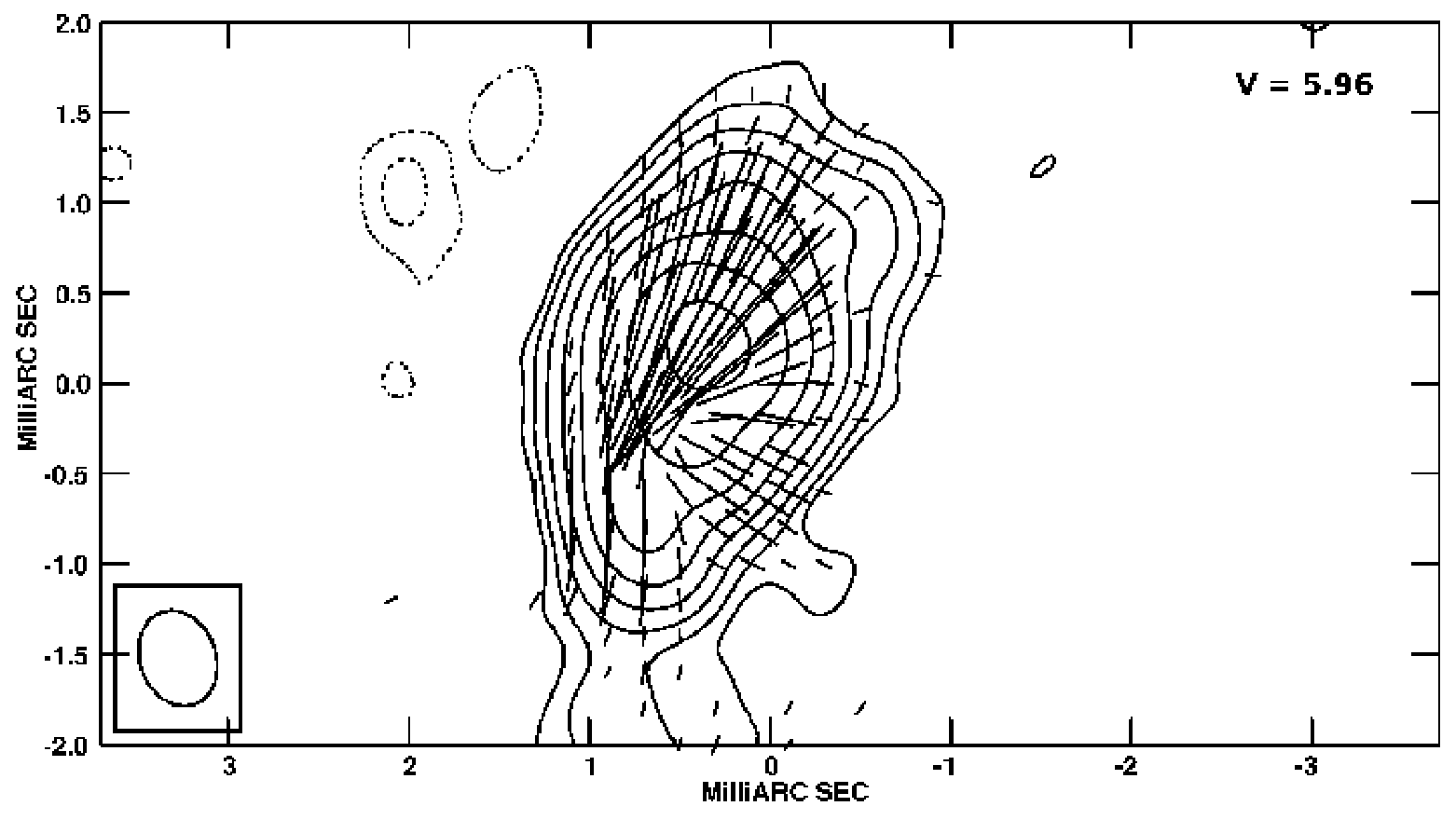} \\
\includegraphics[angle=0,scale=0.25]{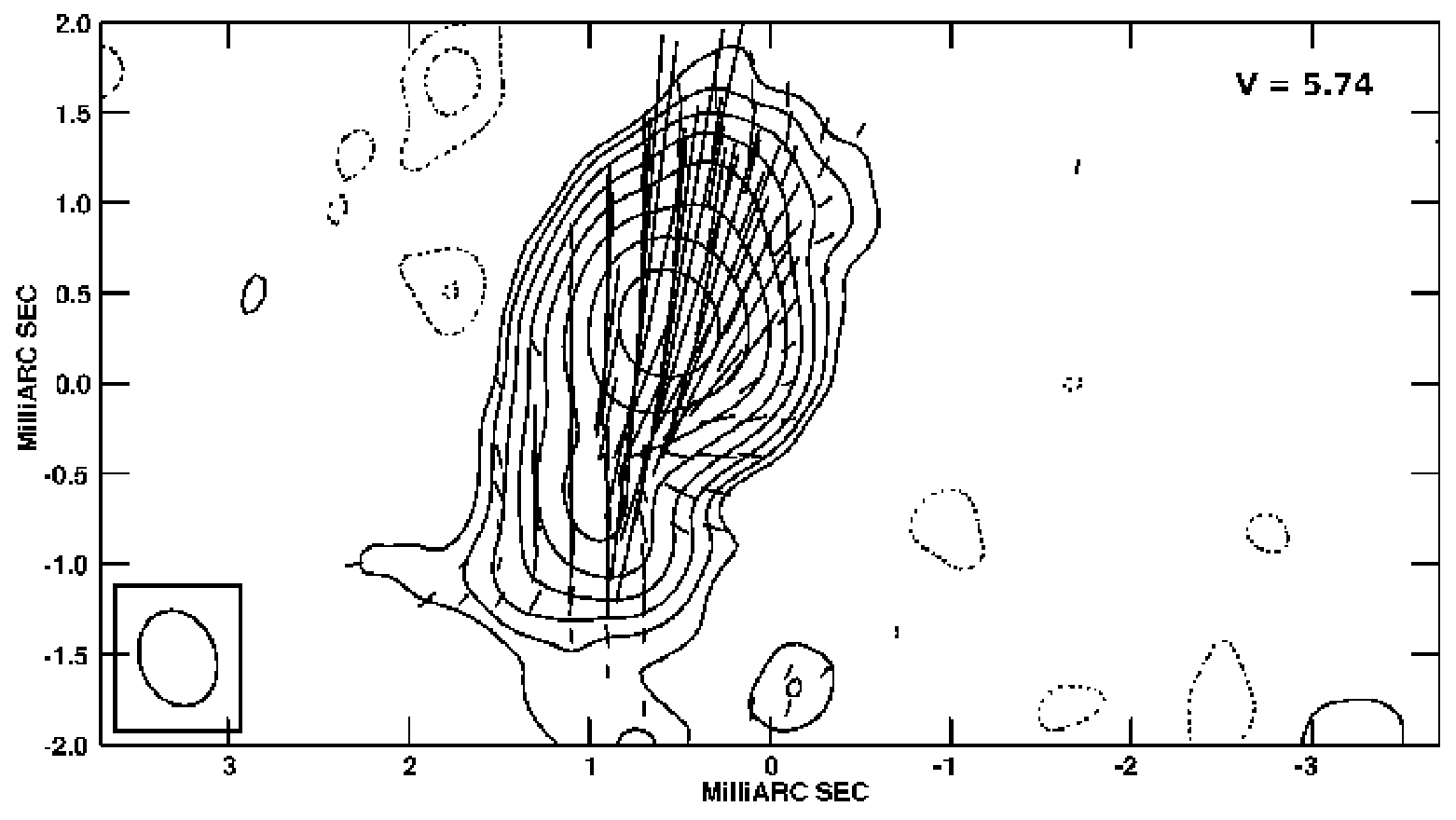} &
\includegraphics[angle=0,scale=0.25]{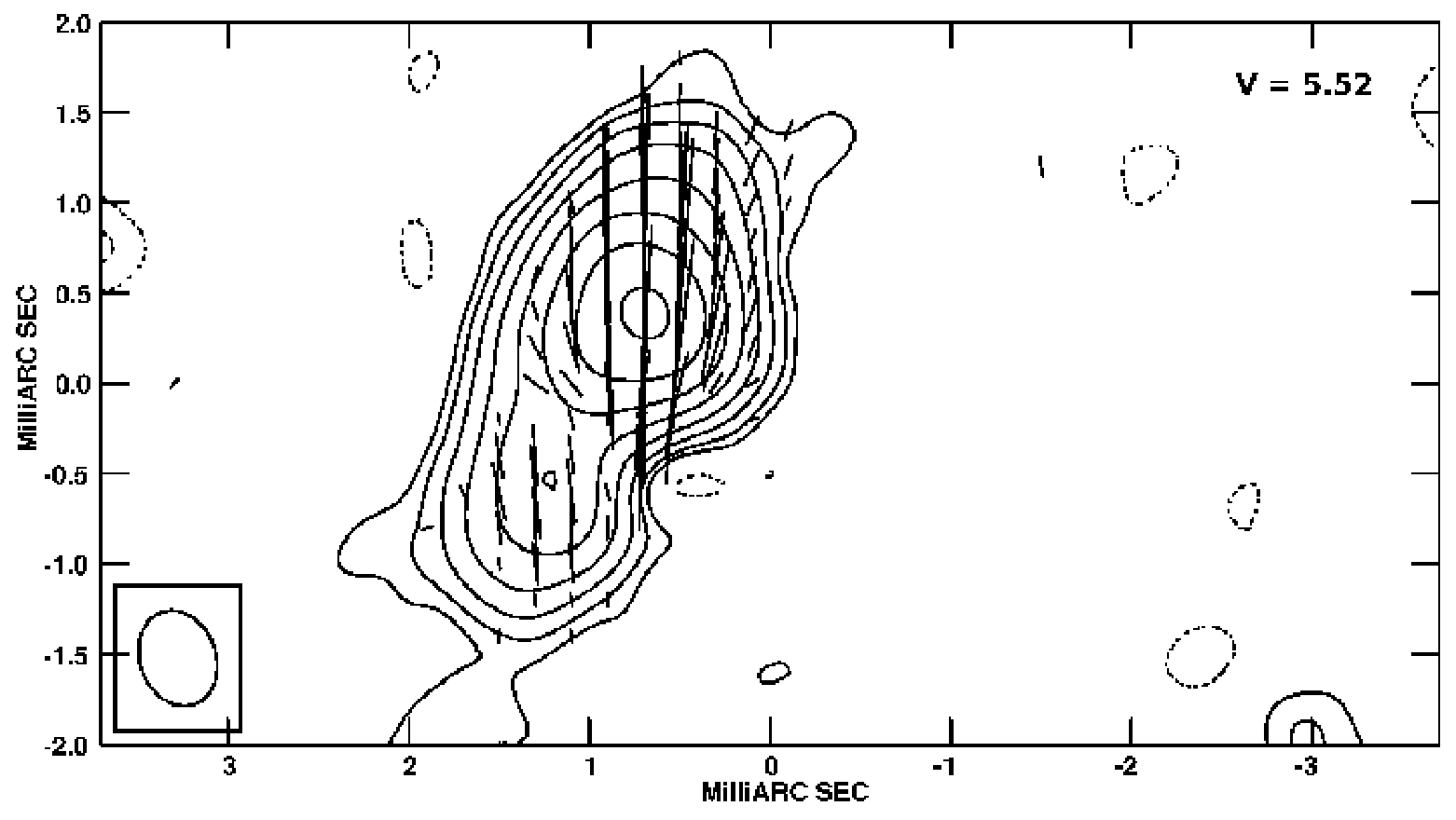} \\
\includegraphics[angle=0,scale=0.25]{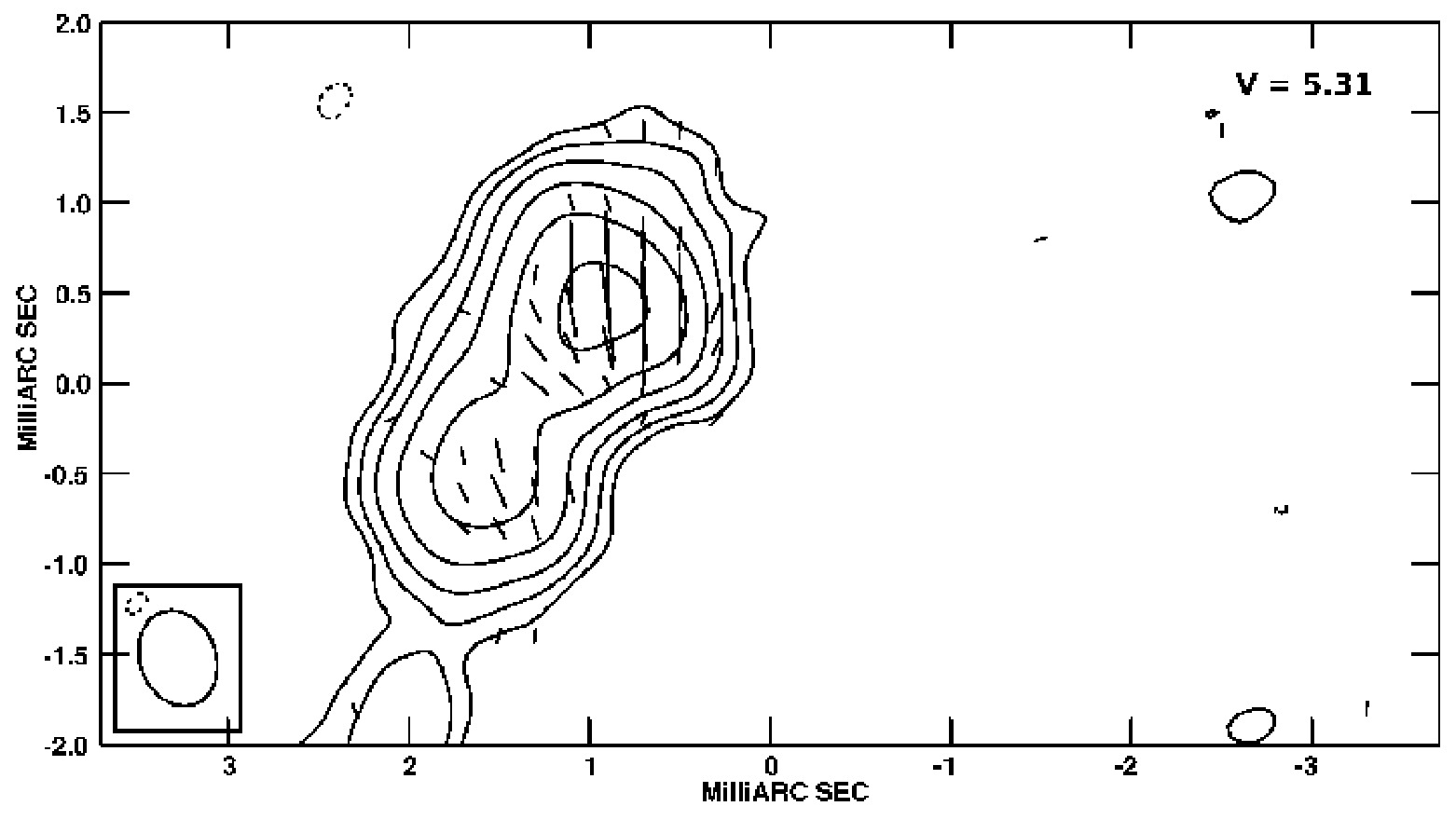} &
\includegraphics[angle=0,scale=0.25]{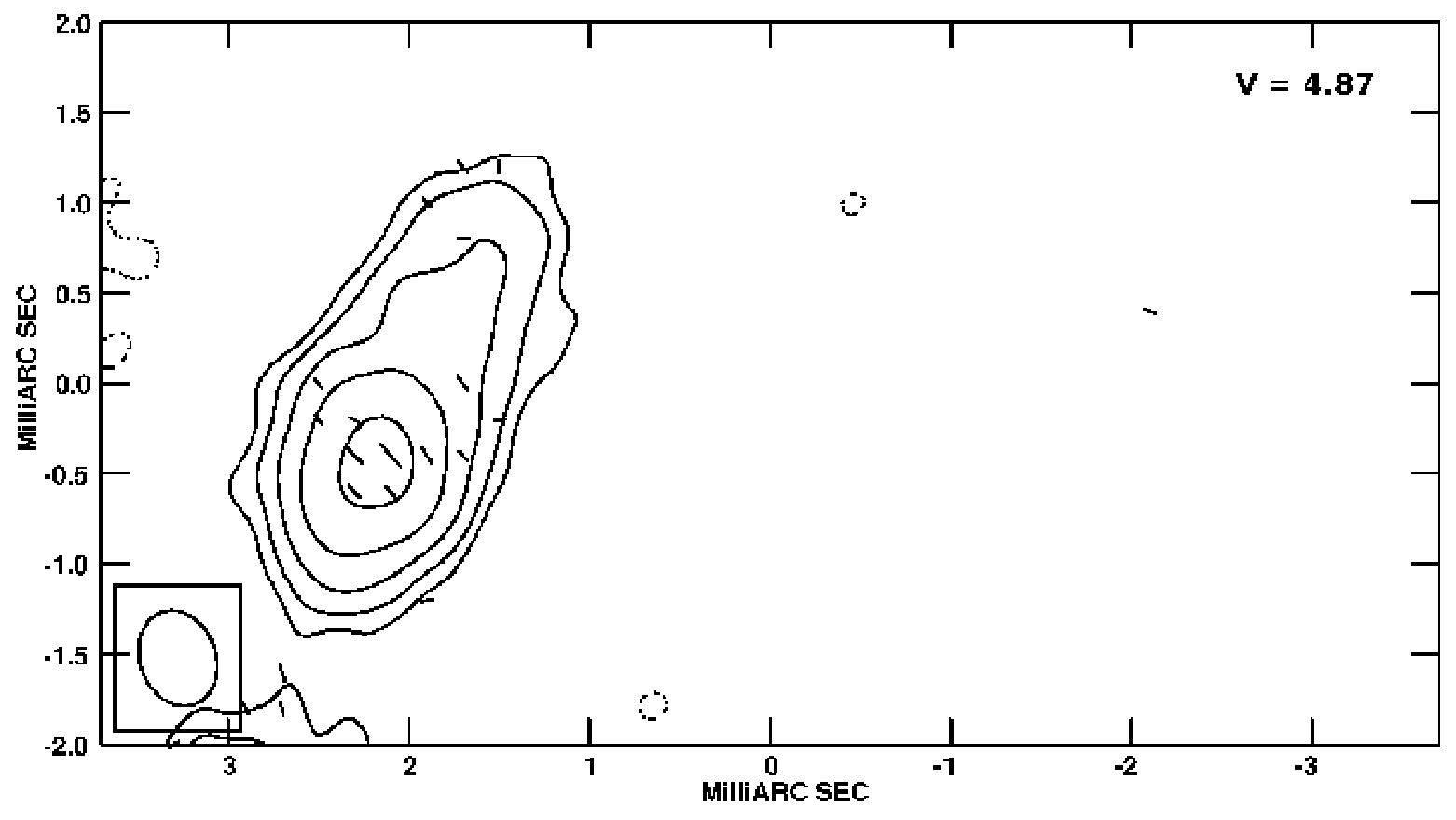} \\
\end{tabular}
\caption{\small Stokes $I$ channel images, labeled with  LSR velocity (km/s), plotted at contour levels $\{-12,-6,-3, 3,6,12,24,48,96,192,384,768\}  \sigma$, where $\sigma$ = 15.7 mJy/beam is the off-source rms in an early frequency channel in this sequence. Overlaid linear polarization vectors are drawn as in Figure~\ref{fig-sqrev-pchi} but with scale such that $P$= 1.05 Jy/beam has length 1 mas.}
\label{fig-reversal-pchi}
\end{figure*}

\begin{figure*}
\centering
\begin{tabular}{cc}
\includegraphics[angle=0,scale=0.25]{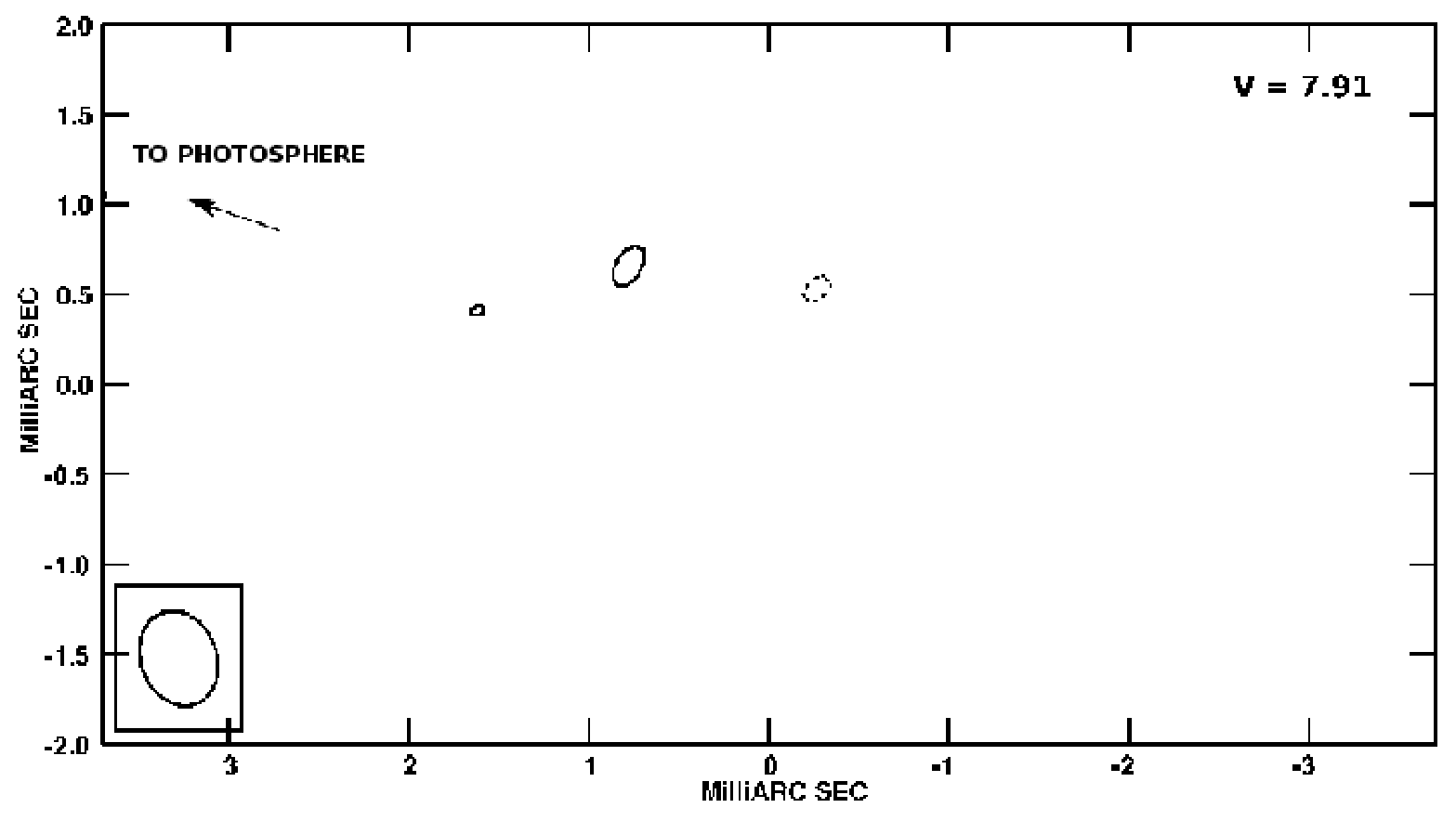} &
\includegraphics[angle=0,scale=0.25]{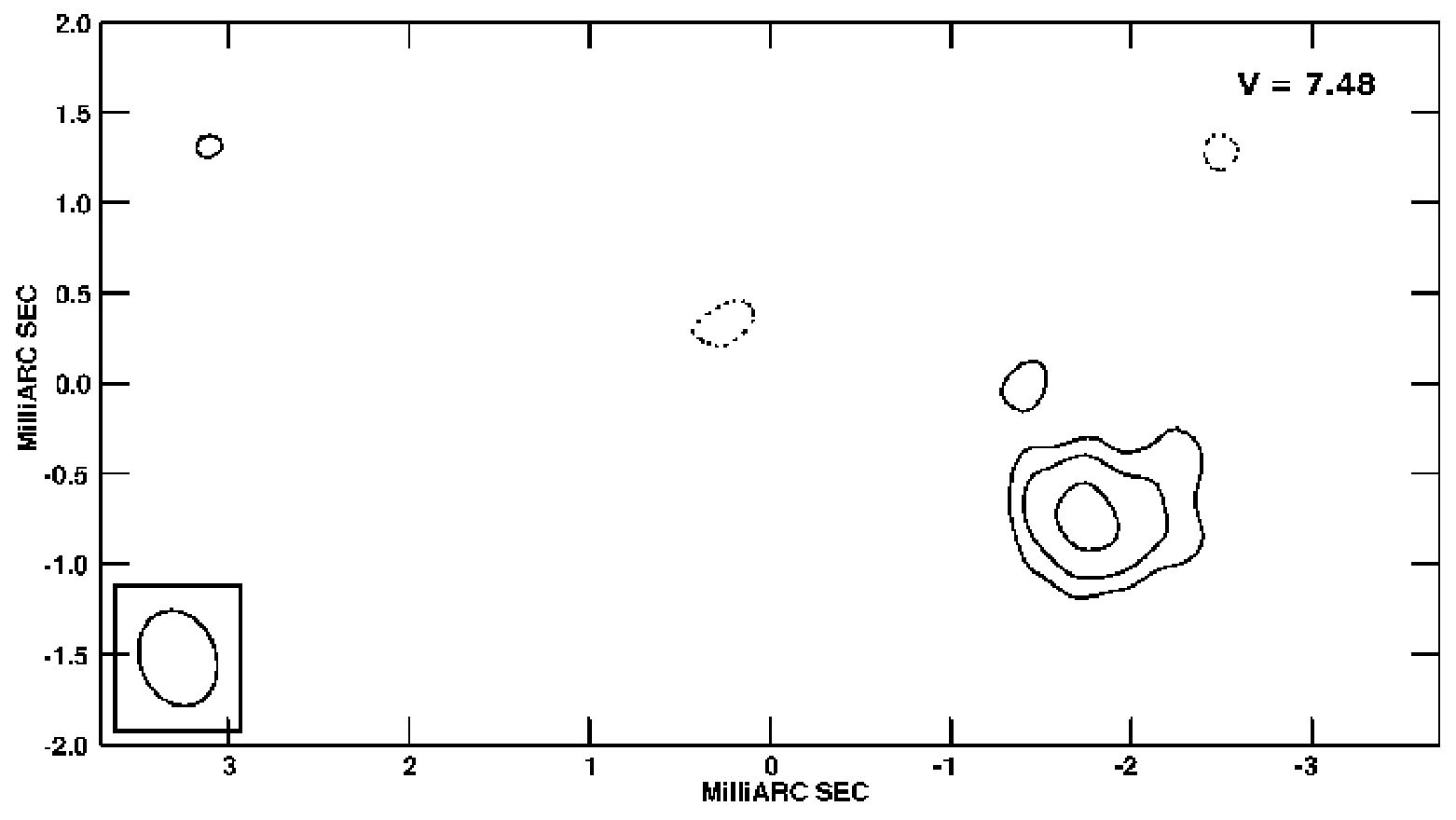} \\
\includegraphics[angle=0,scale=0.25]{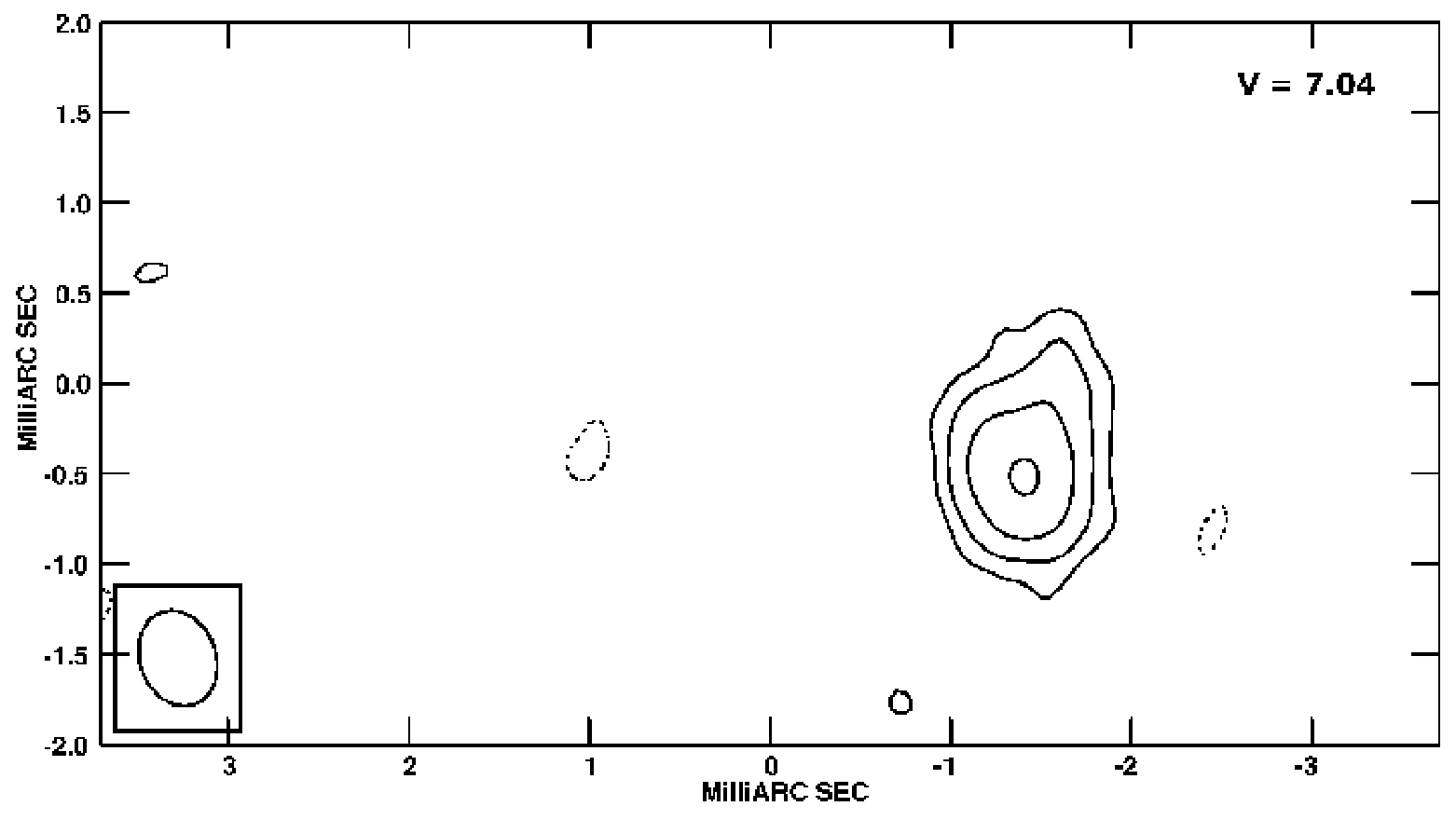} &
\includegraphics[angle=0,scale=0.25]{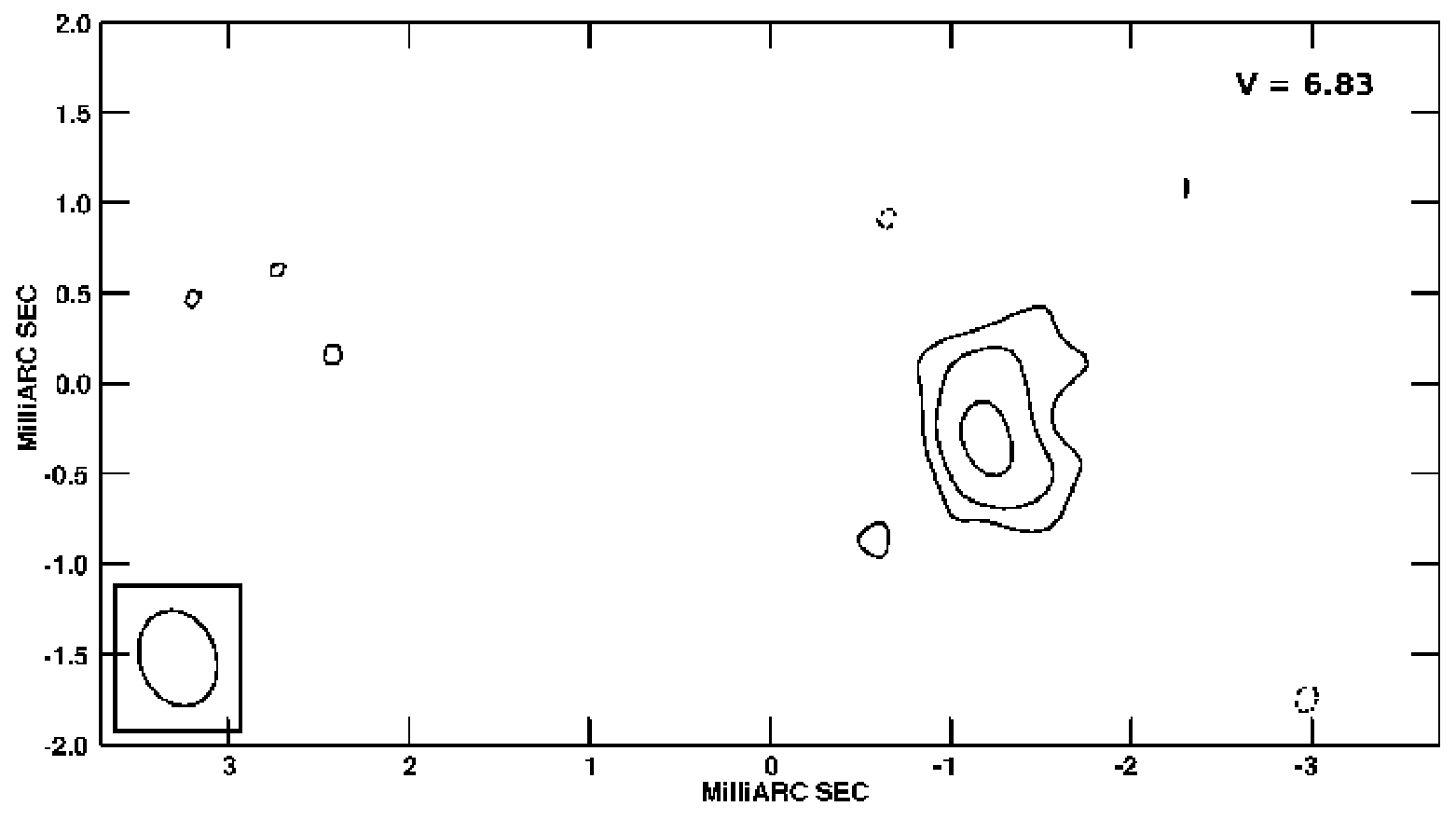} \\
\includegraphics[angle=0,scale=0.25]{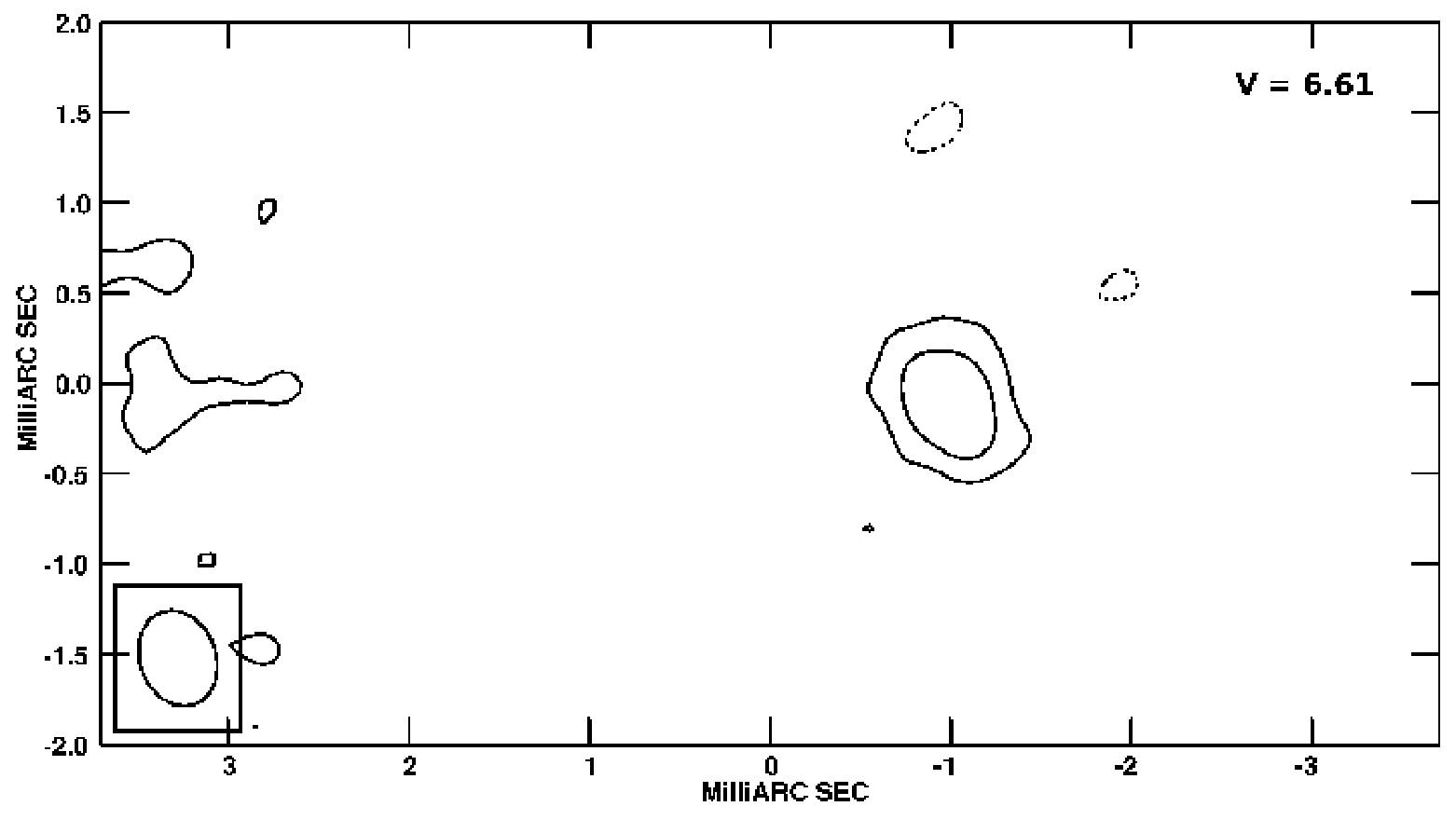} &
\includegraphics[angle=0,scale=0.25]{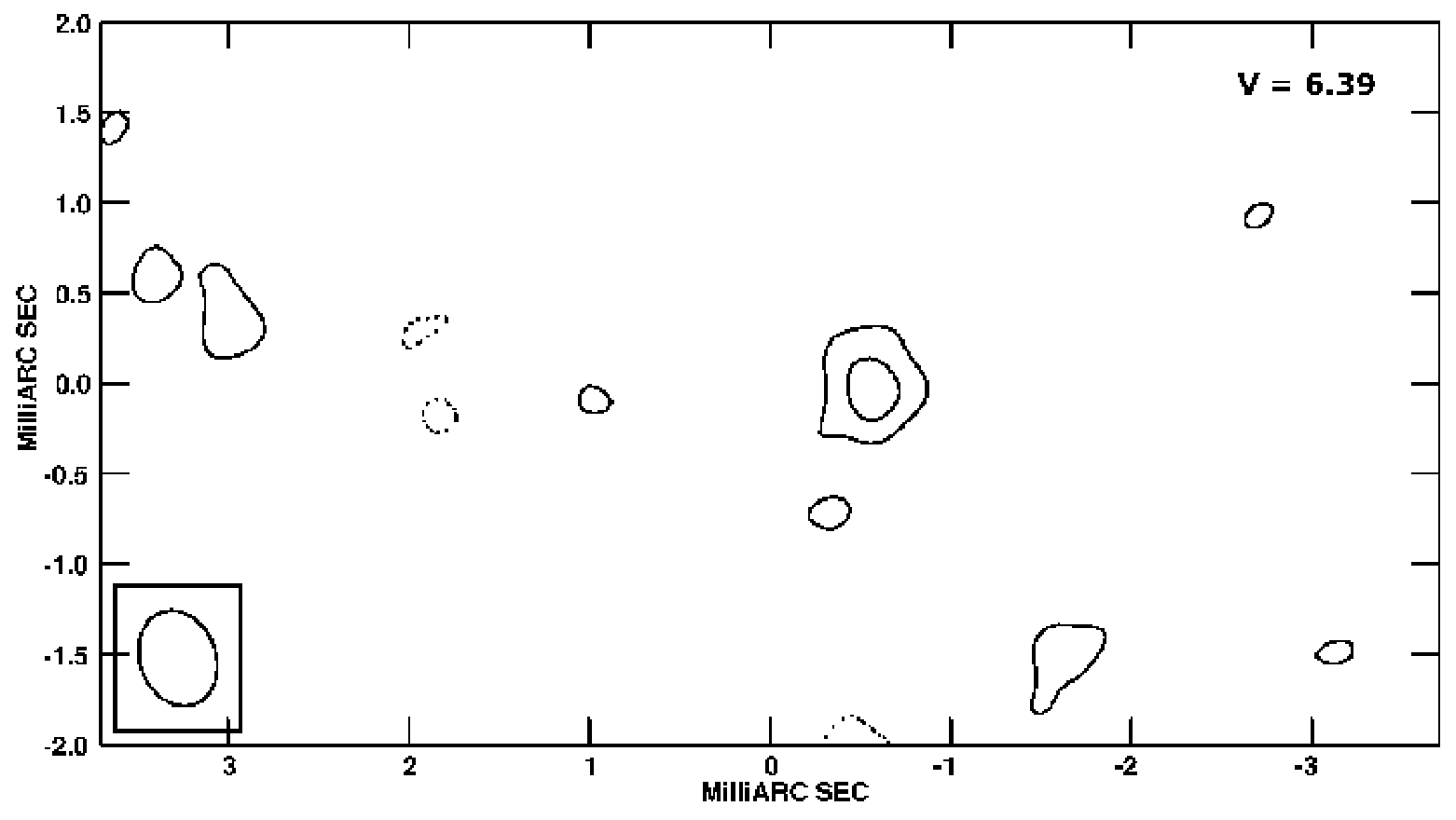} \\
\includegraphics[angle=0,scale=0.25]{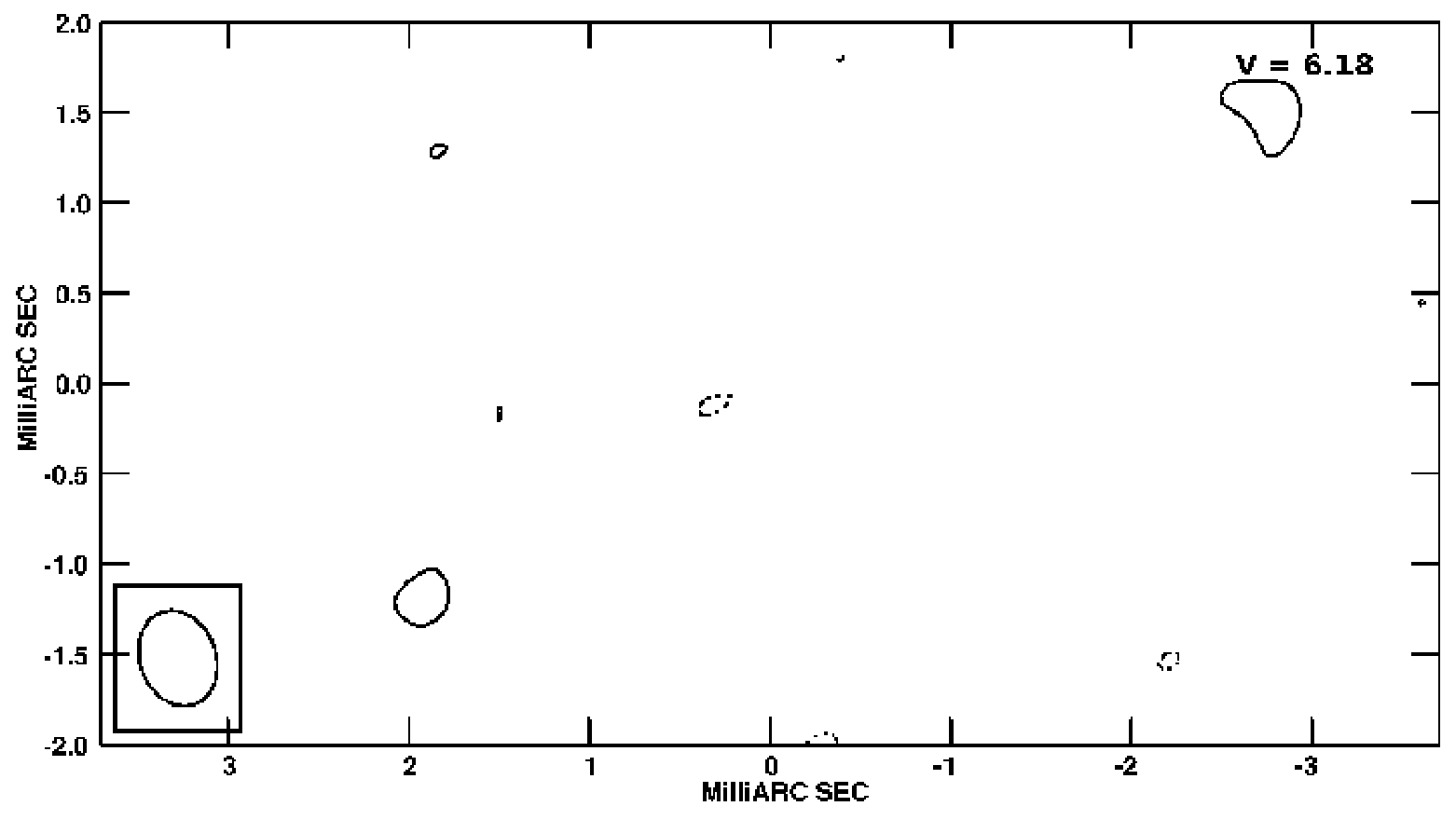} &
\includegraphics[angle=0,scale=0.25]{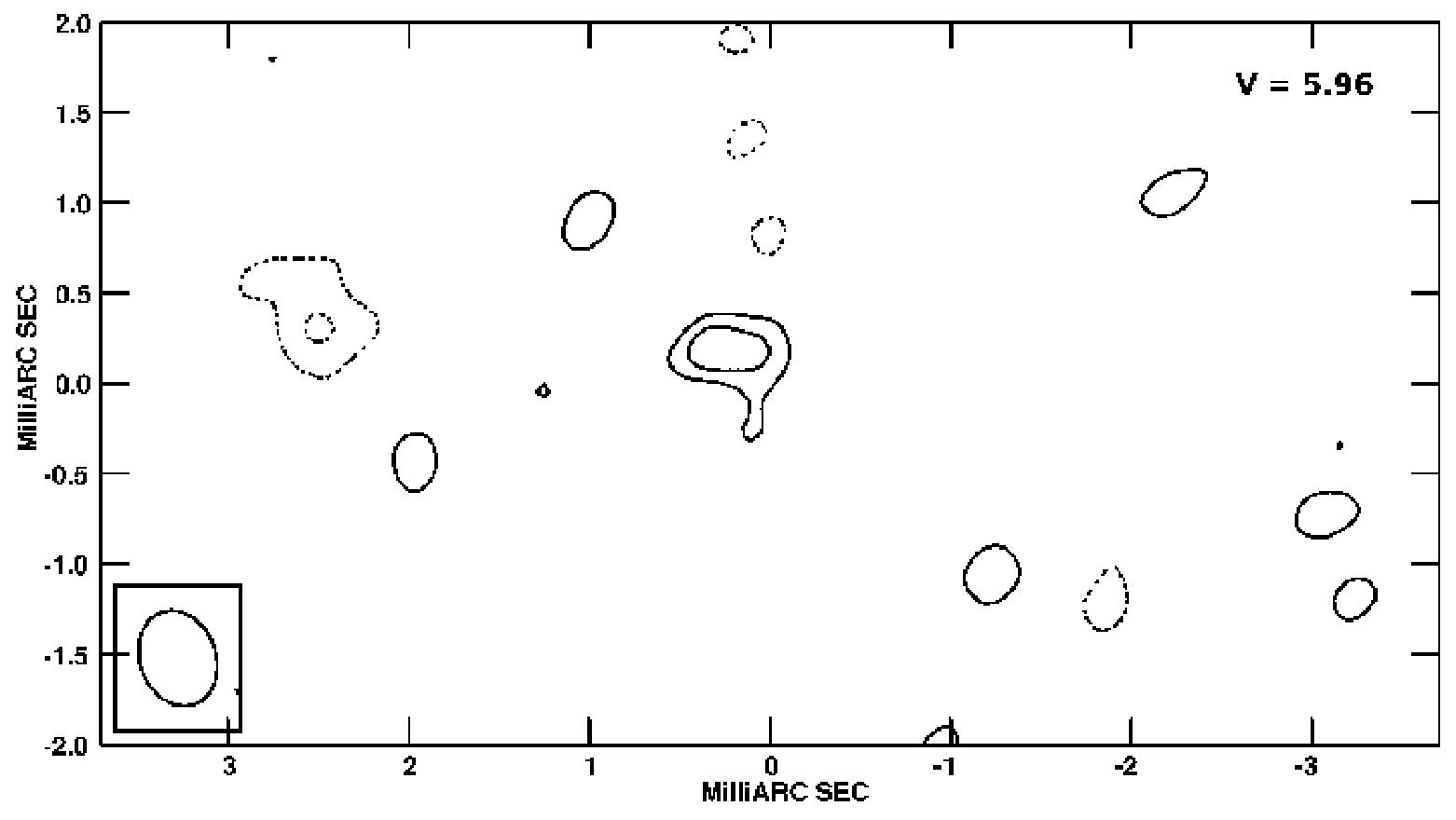} \\
\includegraphics[angle=0,scale=0.25]{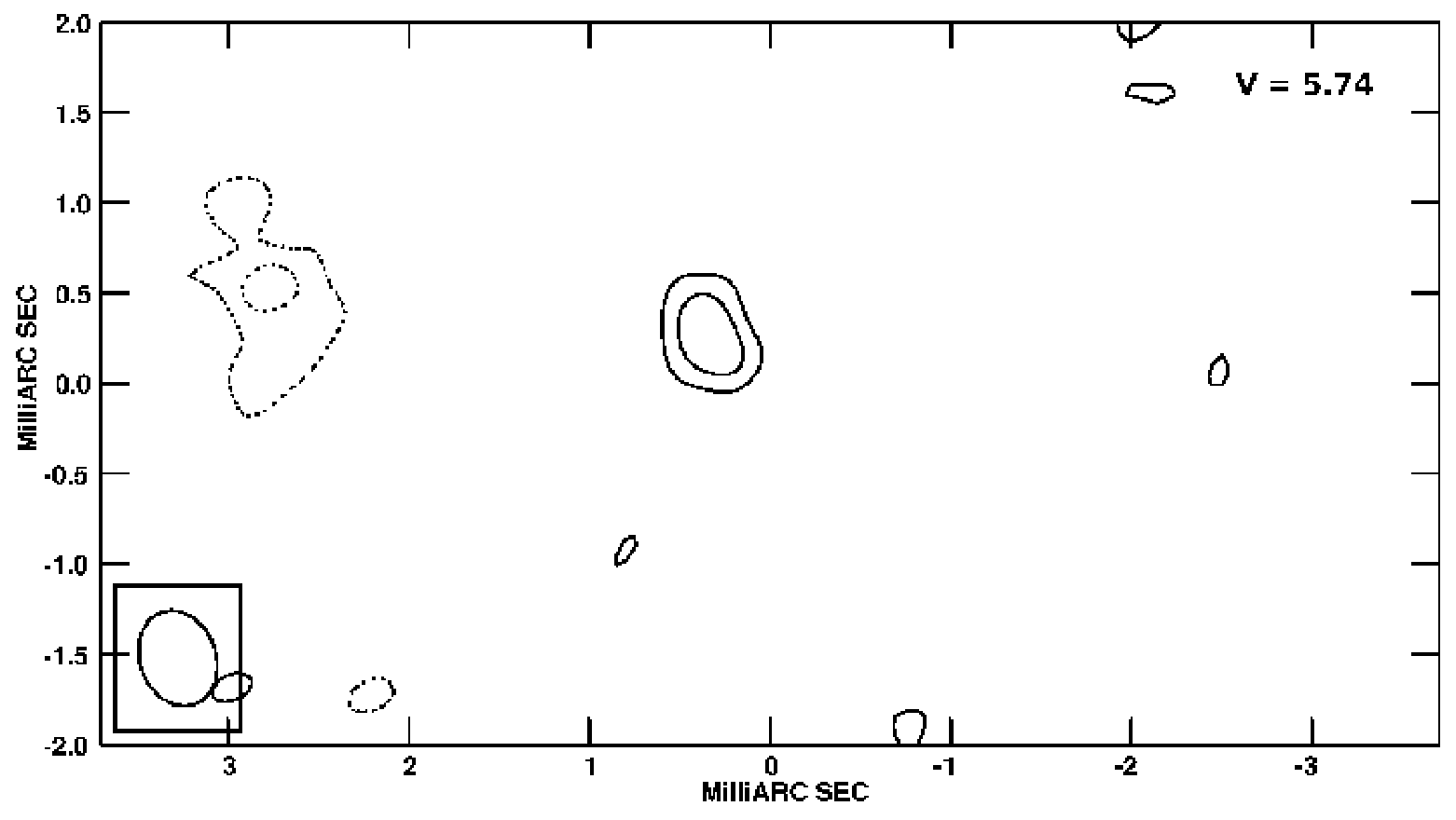} &
\includegraphics[angle=0,scale=0.25]{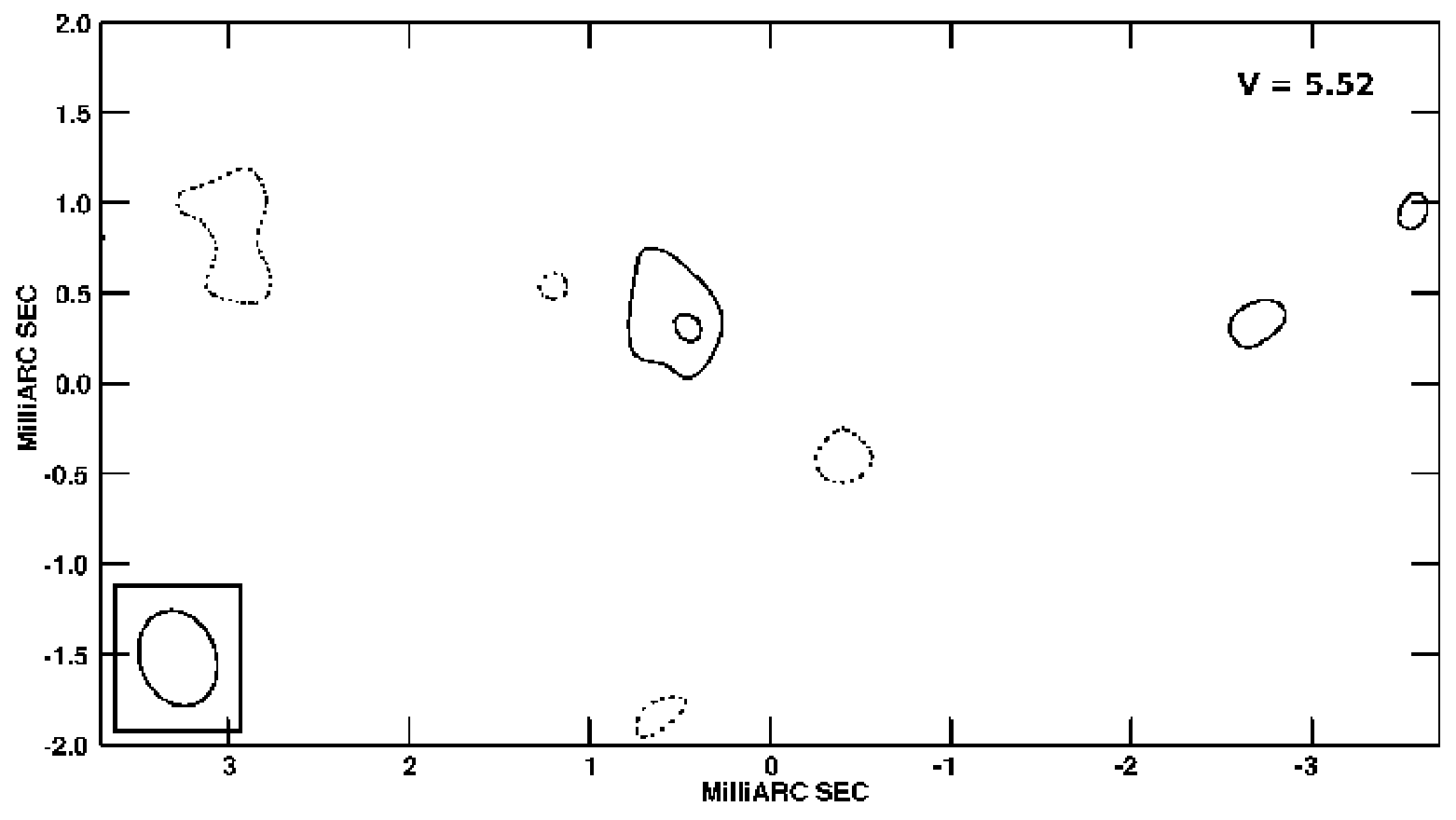} \\
\includegraphics[angle=0,scale=0.25]{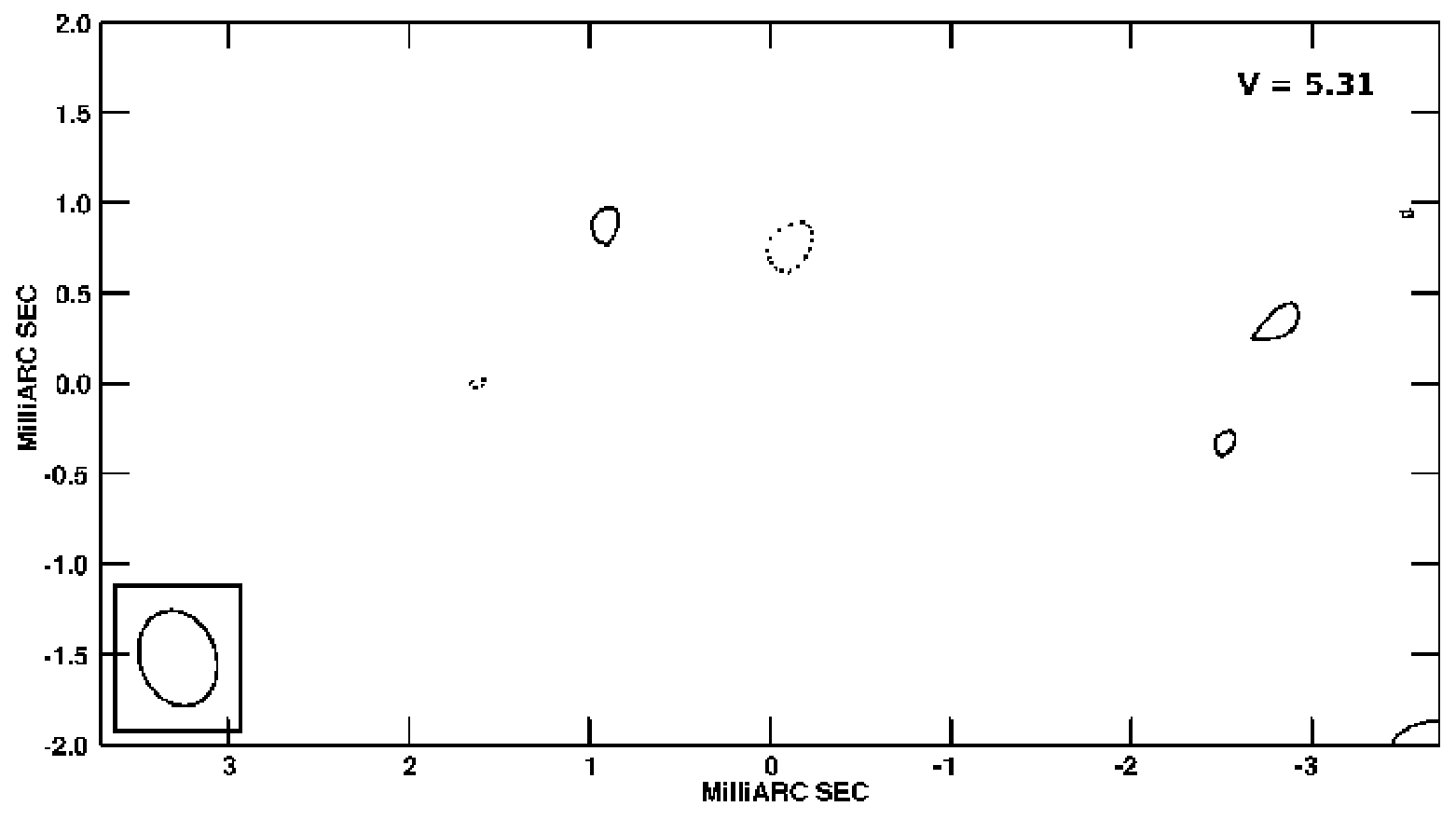} &
\includegraphics[angle=0,scale=0.25]{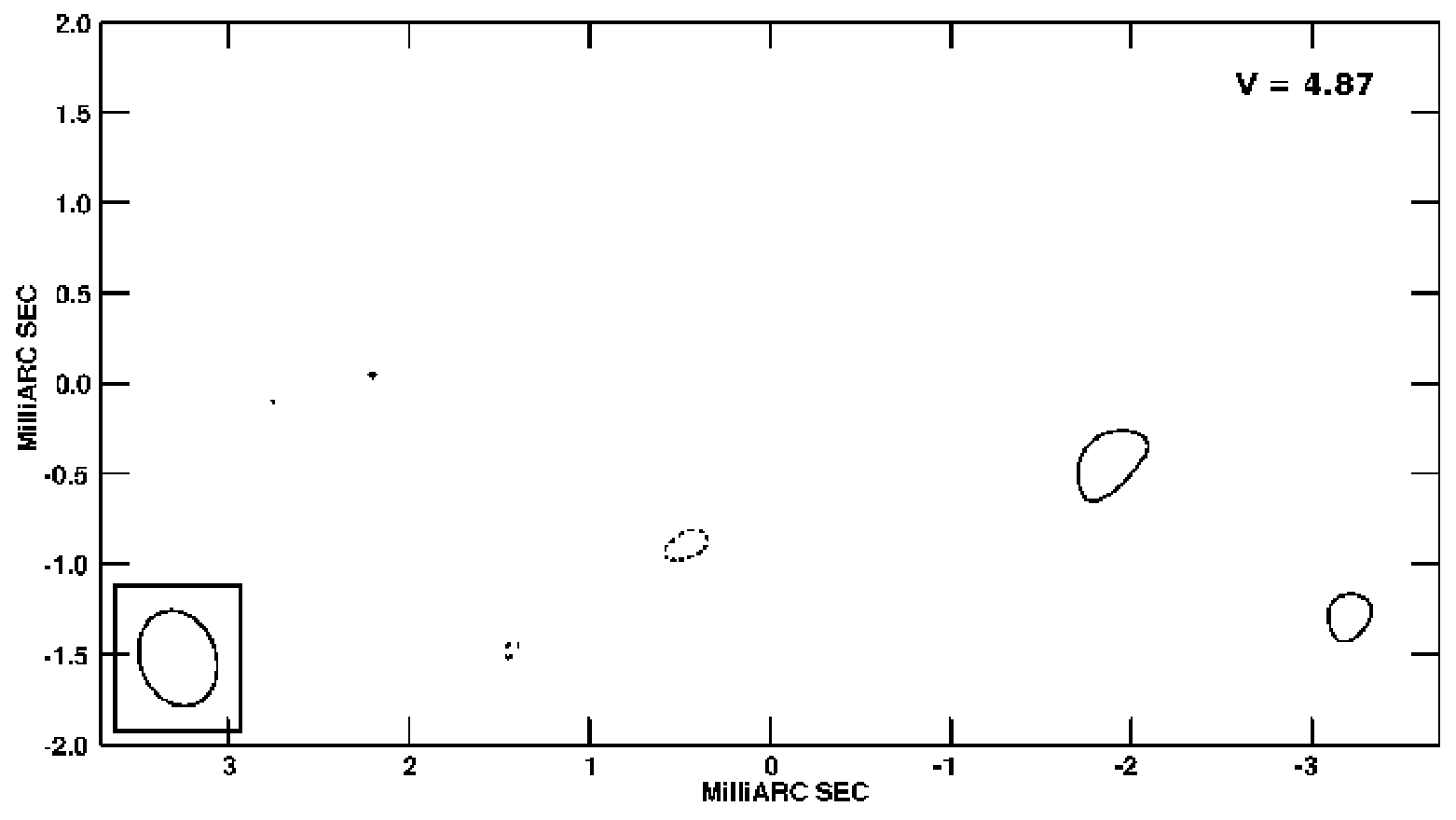} \\
\end{tabular}
\caption{\small Stokes $V$ channel images, matching Figure~\ref{fig-reversal-pchi}, plotted at contour levels $\{-96,-48,-24,-12,-6,-3,3,6,12,24,48,96\}\times \sigma$, where $\sigma$ = 15.7 mJy/beam.}
\label{fig-reversal-v}
\end{figure*}

\begin{figure}
\includegraphics[angle=0,scale=0.4]{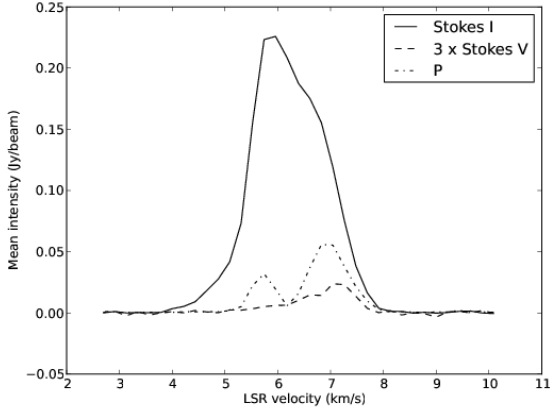}
\caption{Mean-intensity spectrum in Stokes $I$, $V$, and $P=\sqrt{Q^2+U^2}$ over the full image region in Figure~\ref{fig-sqrev-pchi}, enclosing the feature showing the $\frac{\pi}{2}$ rotation in EVPA. This spectrum was obtained by integration over the associated interferometric image cubes. The Stokes $V$ profile is magnified by a factor of three for clarity of presentation.}
\label{fig-spectrum-ivp}
\end{figure}

\begin{figure}
\includegraphics[angle=0,scale=0.4]{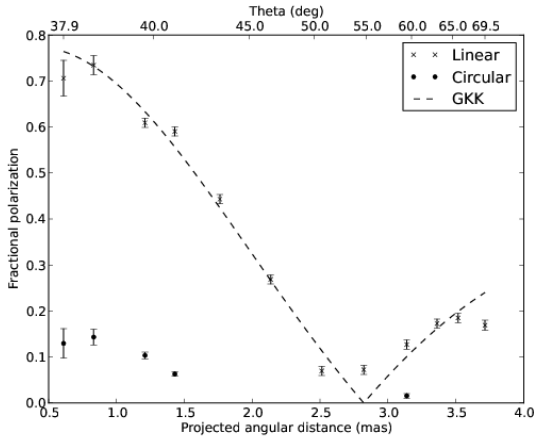}
\caption{Magnitude of fractional linear and circular polarization at the position of the peak Stokes $I$ intensity in each frequency channel spanning the component shown in Figure~\ref{fig-sqrev-pchi}, for all components with an SNR above three. The projected angular distance $d$ is measured as the linear separation from the peak of the upper left channel image in Figure~\ref{fig-reversal-pchi}. The linear polarization model of \citet{gol73}, fitted as described in the text, is plotted as a dashed line. The associated $\theta(d)$ relation is plotted on the upper $x-$axis.}
\label{fig-profile-zplot}
\end{figure}

\begin{figure}
\includegraphics[angle=0,scale=0.4]{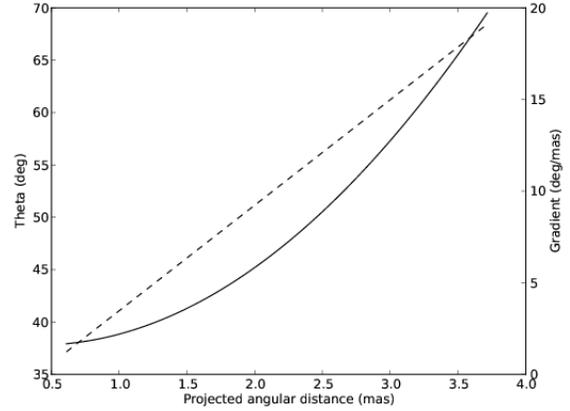}
\caption{Solution for $\theta(d)$ ({\it left axis}) derived in the fit to the GKK model shown in Figure~\ref{fig-profile-zplot}. The gradient of $\theta(d)$ with respect to $d$  is plotted as a dashed line ({\it right axis}).}
\label{fig-profile-chiplot}
\end{figure}
\begin{figure}
\includegraphics[angle=0,scale=0.4]{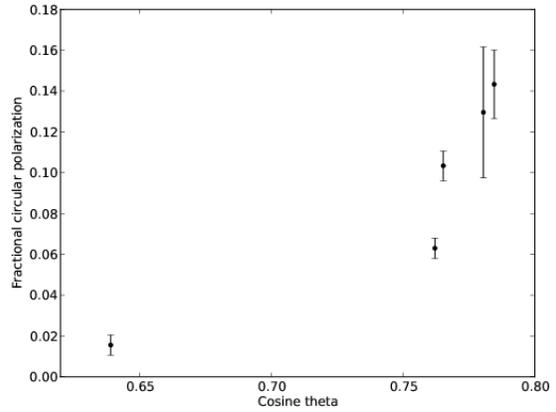}
\caption{Fractional circular polarization as a function of $\cos \theta$, where $\theta$ is the angle between the magnetic field and line of sight, computed according to the \citet{gol73} model, as $\cos \theta=\sqrt{1-\frac{2}{3(m_l+1)}}$.}
\label{fig-profile-mcplot}
\end{figure}

\end{document}